\documentclass[11pt]{article}

\usepackage{amsmath,amssymb,amsfonts,mathtools}
\usepackage{amsthm}
\makeatletter
\let\paper@newtheorem\newtheorem
\def\newtheorem#1{\@ifnextchar[{\paper@skipthm}{\@gobble}}
\def\paper@skipthm[#1]#2{}
\makeatother
\usepackage{jmlr2e}
\makeatletter
\let\newtheorem\paper@newtheorem
\makeatother
\usepackage[T1]{fontenc}
\usepackage[utf8]{inputenc}
\usepackage{lmodern}
\usepackage{microtype}
\usepackage{booktabs,tabularx,array}
\usepackage{longtable}
\usepackage{caption}
\usepackage{algorithm}
\usepackage{algpseudocode}
\usepackage{enumitem}
\usepackage{xcolor}
\usepackage{tikz}
\usetikzlibrary{arrows.meta,positioning,fit}
\usepackage{float}
\usepackage{placeins}
\usepackage{hyphenat}
\usepackage{xurl}

\definecolor{darkblue}{rgb}{0.0,0.0,0.65}
\usepackage[colorlinks,hypertexnames=false]{hyperref}
\hypersetup{breaklinks,linkcolor=darkblue,urlcolor=darkblue,citecolor=darkblue,anchorcolor=darkblue}
\usepackage[capitalise,nameinlink,noabbrev]{cleveref}
\providecommand{\doi}[1]{doi: \href{https://doi.org/#1}{\begingroup\urlstyle{rm}\nolinkurl{#1}\endgroup}}

\theoremstyle{plain}
\newtheorem{theorem}{Theorem}
\newtheorem{proposition}{Proposition}

\newtheorem{lemma}{Lemma}
\theoremstyle{definition}

\newtheorem{assumption}{Assumption}
\theoremstyle{remark}
\newtheorem{remark}{Remark}

\crefname{assumption}{Assumption}{Assumptions}
\crefname{remark}{Remark}{Remarks}
\crefname{appendix}{Appendix}{Appendices}
\crefname{subappendix}{Appendix}{Appendices}

\newcommand{\A}{\mathcal A}
\newcommand{\V}{\mathcal V}
\newcommand{\W}{\mathcal W}
\newcommand{\Rset}{\mathcal R}
\newcommand{\Fset}{\mathcal F}
\newcommand{\Xset}{\mathcal X}
\newcommand{\Dset}{\mathcal D}
\newcommand{\Jset}{\mathcal J}
\newcommand{\E}{\mathbb E}
\newcommand{\Pp}{\mathbb P}
\newcommand{\Nzero}{\mathbb N_0}
\newcommand{\trans}{\mathsf T}
\newcommand{\given}{\,|\,}
\newcommand{\bx}{\boldsymbol x}
\newcommand{\bfv}{\boldsymbol f}
\newcommand{\bz}{\boldsymbol z}
\newcommand{\by}{\boldsymbol y}
\newcommand{\bb}{\boldsymbol b}
\newcommand{\bR}{\boldsymbol R}
\newcommand{\bX}{\boldsymbol X}
\newcommand{\bZ}{\boldsymbol Z}
\newcommand{\bF}{\boldsymbol F}

\newcommand{\diag}{\operatorname{diag}}

\setlist{nosep,leftmargin=2.1em}

\firstpageno{1}

\begin{document}

\title{Conditional Stochastic User Equilibrium\\ as a Gibbs Distribution: Tractable Flow Laws\\ and Exact Path-Free Sampling}

\author{\name Tianyue Qiao \email tianyue.qiao@mail.mcgill.ca
\AND
\name Lijun Sun\thanks{Corresponding author.} \email lijun.sun@mcgill.ca\\
       \addr Department of Civil Engineering\\
       McGill University\\
       817 Sherbrooke Street West, Montr\'eal, Qu\'ebec H3A 0C3, Canada}

\editor{}

\maketitle

\begin{abstract}
Treating traffic counts as data calls for the joint distribution of link flows. Stochastic user equilibrium (SUE) gives a single flow vector, and independent route choice at fixed probabilities ignores congestion feedback between travelers. Conditional stochastic user equilibrium (CSUE), in which each traveler of a finite population chooses a route by logit conditional on the routes of all others, contains this feedback, but computing with it at scale meets three obstacles: the individual state stores every traveler's route and is updated one traveler at a time, aggregated flows carry combinatorial weights that are generally not explicit, and joint updates of many travelers must account for the congestion they impose on one another. Under separable link costs and a common route-choice precision, the CSUE conditionals are those of a unique Gibbs distribution whose energy is a left-endpoint discrete congestion potential, and for continuous nondecreasing costs the law concentrates on deterministic logit SUE in the large-population limit. When the routes of each origin are the paths of an acyclic bush, retaining origin- or OD-level labels gives an unnormalized layered link-flow law with an explicit product-form multiplicity. On this representation we build a block Metropolis--Hastings sampler whose proposal is an integer version of Dial's loading under frozen costs. It leaves the CSUE law on the bush route sets invariant without enumerating paths or storing individual routes, and at stationarity the mean rejection rate of a block lies between one and two times the total variation distance between its proposal and its exact conditional. In numerical experiments, the sampler reproduces reference laws obtained by exact enumeration and by traveler-level Gibbs sampling, whereas independent route choice at frozen costs reproduces mean flows but overstates link-flow standard deviations by up to 48\% in an enumerable example and distorts correlations. On Sioux Falls, the block sampler attains 76.1 times the minimum link-flow effective sample size per CPU second of traveler-level Gibbs sampling, and on Chicago Sketch, with more than one million travelers, its sampled covariance is used to assess traffic-count locations. The CSUE law supplies a latent assignment model with congestion feedback for statistical inference from traffic counts.
\end{abstract}

\begin{keywords}
conditional stochastic user equilibrium, distributional traffic assignment, Gibbs distribution, potential games, path-free sampling, bush-based loading, Markov chain Monte Carlo
\end{keywords}

\section{Introduction}

Traffic assignment allocates fixed origin--destination (OD) demand among routes while accounting for travel costs that depend on congestion. Under Wardrop's deterministic user equilibrium, all routes used by an OD pair have equal and minimal cost, so no traveler can reduce travel cost by changing routes unilaterally \citep{Wardrop1952}. Stochastic user equilibrium (SUE) relaxes perfect cost perception by introducing random errors in perceived route costs. Route choice is therefore probabilistic, and equilibrium requires the aggregate flows implied by these probabilities to reproduce the costs on which the probabilities are based \citep{DaganzoSheffi1977,Sheffi1985}. In its conventional formulation, individual random choices are represented by aggregate route proportions, so the solution remains a single deterministic flow vector. This point representation is useful for describing average network conditions, but it does not characterize the distribution of realized path and link flows generated by a finite number of stochastic route choices.

That distribution is needed whenever traffic counts are treated as data. To predict the flow on an unobserved link from counts elsewhere, or to decide where the next counter should go, one needs not just the means of link flows but their covariances. A common way to obtain them is to let travelers choose routes independently, with route probabilities fixed in advance, which gives Poisson or multinomial route flows \citep[e.g.,][]{Hazelton2000}. Congestion breaks this independence. A traveler who joins a route raises its cost for the others, some of whom respond by leaving it, so a fluctuation is damped on the route where it arises and passed on to the alternatives. In the small example of \cref{subsec:exp-validation}, where the exact law can be enumerated, independent route choice at frozen costs reproduces mean link flows to within 0.6\% but overstates link-flow standard deviations by up to 48\% and distorts the correlations between links. We refer to models whose output is a probability law over network states, rather than a single state, as \emph{distributional traffic assignment}.

\citet{Hazelton1998} introduced conditional stochastic user equilibrium (CSUE) to represent this distribution of realized network states directly. For each labeled traveler, CSUE specifies a route choice conditional on the routes selected by all other travelers, with candidate route costs evaluated after removing that traveler's own contribution from the network. The corresponding equilibrium object is a joint probability law over complete labeled route profiles. Aggregate path flows and total link flows are deterministic functions of the route profile and inherit pushforward distributions. Because each traveler responds to the realized routes of the others, the law contains the congestion feedback that independent route choice leaves out. \citet{Hazelton1998} also interpreted CSUE as the long-run limit of a continuous-time assignment process and proved that it converges to deterministic SUE as demand grows.

Three computational obstacles have limited the use of CSUE as a model of flow variability. First, the labeled state stores every traveler's route, and the Gibbs sampler implied by the full conditionals updates one traveler at a time, so a sweep costs one route-choice update per traveler. Second, aggregation changes the induced law. Mapping labeled route profiles to path counts introduces a multinomial multiplicity, because many profiles can produce the same path-count vector. Mapping path counts further to total link flows requires summing these multiplicities over all feasible multicommodity decompositions of each link flow, so an aggregate state is useful for sampling only if its combinatorial weight stays explicit. Third, updating many travelers jointly must account for the congestion they impose on one another: loading a group of travelers independently at frozen costs is cheap, but it does not sample their conditional law. Existing computational treatments of equilibrium flow distributions either simulate the underlying assignment process \citep{HazeltonLeePolak1996,Watling1996}, which is costly on large networks, or rely on Gaussian and moment approximations (\cref{sec:literature}), whose accuracy for small counts and tail probabilities can be difficult to assess.

The main contribution of this paper is a state representation and a sampler designed for each other. On fixed origin bushes, finite-population logit CSUE is written exactly as an unnormalized law over layered link flows with computable combinatorial weights, and this law is sampled by a Markov chain Monte Carlo (MCMC) method whose block Metropolis--Hastings (MH) updates use an integer version of Dial's loading as proposal, without enumerating paths or storing individual routes. The construction rests on two known results. First, the CSUE conditionals are the logit revision rule of an atomic congestion game, so the joint law is a Gibbs distribution in the potential of that game \citep{Blume1993}. Second, when the routes from each origin are the paths of an acyclic subnetwork (a bush), the number of labeled route assignments that produce a given link-flow pattern is a product of multinomial coefficients over nodes \citep{Akamatsu1997}. Together they give an explicit unnormalized joint law for the link flows of the individual origins or OD pairs, which we call layers. Paths are summed out within each layer, and the remaining sum over decompositions of total link flow into layers is never evaluated, because the sampler keeps the layers as its state. The proposal carries the same multiplicities, so they cancel from the acceptance ratio. In more detail, the paper makes three contributions:

\begin{enumerate}[label=(\arabic*)]
  \item \emph{Flow laws.} Under separable link costs and a common route-choice precision, we give the joint law of logit CSUE and the laws it induces on path counts, on origin- or OD-layered link flows, and on total link flows. The energy of the Gibbs law is Rosenthal's potential for link costs shifted to match the leave-one-out convention of CSUE, and it is a left-endpoint Riemann sum for the Beckmann potential. Aggregation to path counts adds a multinomial multiplicity. For continuous nondecreasing costs the path-count law concentrates on deterministic logit SUE as individual flow contributions vanish at fixed physical demand, in agreement with \citet{Hazelton1998} and with large-deviation results for logit choice in potential games \citep{SandholmStaudigl2018}.
   On bushes, the multiplicity of a layer is a product of node-level multinomial coefficients. We organize this counting, which goes back to \citet{Akamatsu1997}, into the layered target that the sampler uses, so that origin, OD, and labeled sub-batch updates share one exact representation.

  \item \emph{Sampler.} We develop a path-free block MH sampler on layered link flows. Its proposal is an integer sampling version of Dial's loading under frozen costs and has an explicit law, so the acceptance ratio reduces to a potential difference and a linear term. The sampler leaves the CSUE law on the fixed bush route sets invariant. With one block per origin, the arithmetic work of a sweep depends on the sizes of the bushes rather than on the number of travelers; a block whose proposals are rejected too often can be split, at the cost of more updates per sweep. At stationarity, the mean rejection rate of a block lies between one and two times the total variation distance between its proposal and its exact conditional. Averaged over backgrounds, the rejection rate is thus an observable measure of the conditional error that the same update would make without the accept--reject step.
  \item \emph{Numerical evidence.} Where a reference law is available, by exact enumeration or by traveler-level Gibbs sampling, the sampler reproduces it, while the same update without the accept--reject step matches mean flows but not standard deviations or correlations. On Sioux Falls we examine how link-flow variability changes with route-choice precision and flow quantum. On Chicago Sketch, with more than a million travelers, we use the sampled covariance to evaluate sets of count locations.
\end{enumerate}%

We treat the forward problem: OD demand, link-cost functions, bush route sets, and the route-choice precision are given, and the task is to characterize and sample the law of network states that they imply. Randomness comes from route choice alone, and day-to-day variation in demand is not represented. What the CSUE law adds to SUE is the dependence between link flows and an exact treatment of links and OD pairs that carry few travelers. The forward law is also the latent assignment model that Bayesian inference of route-choice precision, link-cost parameters, and OD demand from traffic counts would require; \cref{sec:discussion} discusses this extension.

The remainder of this paper is organized as follows. \Cref{sec:literature} reviews related work. \Cref{sec:model} defines the finite-population model and its joint Gibbs law. \Cref{sec:pushforwards} derives the path-, layered-, and link-flow laws and the deterministic SUE limit. \Cref{sec:sampling} presents traveler-level Gibbs sampling and develops the path-free block sampler. \Cref{sec:experiments} reports numerical experiments, and \cref{sec:discussion} concludes with limitations and extensions. Proofs and supplementary results, implementation details, and a notation summary are given in \cref{app:proofs,app:implementation,app:notation}.

\section{Related Literature}
\label{sec:literature}

\paragraph{Distributions of network flows.}
Two broad approaches describe flow variability beyond deterministic SUE. The first models the day-to-day evolution of route choices as a stochastic process and studies its stationary distribution \citep{Cascetta1989,CantarellaCascetta1995}. CSUE arises in the same way, as the long-run law of a process in which travelers revise routes one at a time \citep{Hazelton1998}. Such stationary distributions can be studied by simulating the process \citep{HazeltonLeePolak1996}, although convergence can be very slow when the stationary distribution has several peaks \citep{Watling1996}. \citet{DavisNihan1993} showed that,  as the number of travelers grows, such a process is approximated by a nonlinear deterministic process plus a linear Gaussian fluctuation, and \citet{HazeltonWatling2004} used this type of approximation to compute equilibrium distributions of Markov assignment models, taking the mean from a conventional assignment and the covariance from linearization. Later work synthesized the theory of these processes and their moment approximations \citep{WatlingCantarella2013,WatlingHazelton2018}. \citet{Hazelton2022} showed that, for discrete traffic, deterministic SUE generally cannot coincide with the stationary mean of a time-homogeneous process. The second approach imposes consistency directly on random flows. \citet{Watling2002} equilibrated the moments of the joint flow distribution up to a chosen order and recovered SUE at order one. \citet{ClarkWatling2005} and \citet{NakayamaWatling2014} derived distributions of route flows, link flows, and travel times under stochastic demand and route choice.

\paragraph{Potential games and logit dynamics.}
Atomic congestion games admit an exact potential \citep{Rosenthal1973IJGT,Rosenthal1973,MondererShapley1996}, and the logit (log-linear) revision rule in a potential game has a Gibbs stationary distribution in the potential \citep{Blume1993}. Stochastic stability and large-population limits of such dynamics are well developed \citep{Sandholm2010}, including explicit large-deviation rates for logit choice in potential games \citep{SandholmStaudigl2018}. \Cref{thm:gibbs,thm:large-population-sue} are network instances of these results, and \cref{rem:logit-dynamics} makes the correspondence explicit. \citet{Kleer2023} studied sampling from this Gibbs distribution, showing that single-player logit dynamics can mix exponentially slowly and that, on single-OD extension-parallel networks, path counts with multinomial weights can be sampled in polynomial time. That analysis gives stronger mixing guarantees on a more restricted network structure. The present paper treats several OD pairs on fixed origin bushes and provides a path-free layered block construction for them, without a general fast-mixing guarantee.

\paragraph{Path-free network loading.}
\citet{Dial1971} showed that logit loading over a suitably restricted path set can be carried out by node recursions without enumerating paths. Subsequent work interpreted these recursions as Markov processes, extended them to more general path sets, and used the Markov property to decompose path-choice entropy into link-flow terms \citep{Bell1995,Akamatsu1996,Akamatsu1997}. \citet[Section~1.2]{Akamatsu1997} gave a factorial count of labeled assignments producing a single-OD link-flow pattern, before applying Stirling's approximation. Section~4.2 of that paper uses destination-independent backward choice probabilities to aggregate flows by origin. Related link-based route choice and equilibrium models avoid choice-set enumeration altogether \citep{BaillonCominetti2008,FosgerauFrejingerKarlstrom2013}. Origin-based and bush-based algorithms exploit acyclic subnetworks to solve deterministic equilibrium problems without storing paths \citep{BarGera2002,Dial2006,Nie2010}.

\paragraph{Statistical inference on traffic networks.}
Bayesian estimation of OD matrices from link counts has a long history \citep{Maher1983,TebaldiWest1998,Hazelton2000,Hazelton2008}. Route-choice probabilities are usually taken as given in these models, so that travelers choose independently. MCMC inference can require sampling latent route flows consistent with the observed link counts, a set that is usually too large to enumerate \citep{ParryHazelton2013}, and samplers on this set can become trapped unless the polytope of feasible route flows is suitably represented \citep{Hazelton2015}. Recent work also uses traffic assignment models to inform Bayesian OD inference \citep{HazeltonNajim2024}. Where counts are collected is itself a design problem \citep{YangZhou1998,GentiliMirchandani2012}, and variance-reduction criteria for sensor placement are well developed for Gaussian models \citep{KrauseSinghGuestrin2008}.

\section{Finite-Population Logit CSUE and Its Gibbs Distribution}
\label{sec:model}

\subsection{Network, travelers, and route costs}

Let $G=(\V,\A)$ be a finite directed network, and let $\W$ be the set of OD pairs. For $w=(o_w,d_w)\in\W$, let $n_w\in\mathbb N$ be the fixed number of travelers and let $\Rset_w$ be a finite, nonempty set of $o_w$--$d_w$ routes. The route sets are model primitives; \cref{subsec:layered} specifies them as the paths of an origin-rooted acyclic subnetwork, which \cref{sec:sampling} holds fixed during sampling. The labeled travelers of OD pair $w$ form the set $\Jset_w=\{1,\ldots,n_w\}$, and total demand is $N=\sum_{w\in\W}n_w$. Traveler $(w,j)$ chooses a route $R_{wj}\in\Rset_w$, and the complete labeled route profile $\bR=(R_{wj}:w\in\W,\;j\in\Jset_w)$ takes values in the product space $\mathcal S:=\prod_{w\in\W}\Rset_w^{\,n_w}$.

Routes traverse each link at most once. Let $\Delta$ be the link--route incidence matrix, whose column $\Delta_r\in\{0,1\}^{|\A|}$ indicates the links of route $r$. Each traveler contributes one unit to every link on the chosen route, so the induced integer link flow is
\begin{equation}
  \bx(\bR)=\sum_{w\in\W}\sum_{j=1}^{n_w}\Delta_{R_{wj}}.
\end{equation}

Link costs are separable: each link $a\in\A$ has a finite cost function $c_a$ that depends only on the flow on that link, and $\mathbf c(\bx):=(c_a(x_a))_{a\in\A}$. In the finite-population model, costs are evaluated at integer counts, and the cost of route $r\in\Rset_w$ at link flow $\bx$ is
\begin{equation}
  C_{wr}(\bx)=\Delta_r^{\trans}\mathbf c(\bx)=\sum_{a\in r}c_a(x_a).
\end{equation}

\begin{assumption}[Core finite-population setting]
\label{ass:core}
Unless stated otherwise:
\begin{enumerate}[label=(\roman*)]
  \item every traveler has unit mass;
  \item all travelers share a common logit precision $\theta>0$;
  \item link costs are separable and finite on $[0,\infty)$.
\end{enumerate}
\end{assumption}

\subsection{Conditional stochastic user behavior}

Let $\bx^{-wj}:=\bx(\bR)-\Delta_{R_{wj}}$ be the link flow generated by all travelers except $(w,j)$. With deterministic route utilities $-C_{wr}(\bx^{-wj})$ and independent additive Gumbel errors with scale $1/\theta$, utility maximization gives the CSUE full conditional
\begin{equation}
  \Pp\!\left(R_{wj}=r\given\bR_{-(wj)}\right)
  =
  \frac{\exp\{-\theta C_{wr}(\bx^{-wj})\}}
       {\sum_{s\in\Rset_w}\exp\{-\theta C_{ws}(\bx^{-wj})\}},
  \qquad r\in\Rset_w.
  \label{eq:csue-conditional}
\end{equation}
The traveler's own contribution is excluded from every candidate cost, so all routes are evaluated against the common background flow of the other travelers \citep{Hazelton1998}. Here ``conditional'' refers to the other travelers' routes; OD demands and the remaining model inputs are fixed.

A family of full conditionals need not correspond to any joint distribution \citep{Besag1974,ArnoldPress1989}. The conditionals in Equation~\eqref{eq:csue-conditional} are compatible, however, because they are generated by a discrete congestion potential, which we construct next.

\subsection{The left-endpoint discrete congestion potential}

For each link $a$ and integer flow $k\ge0$, define the left-endpoint discrete link potential
\begin{equation}
  \phi_a(k)=\sum_{m=0}^{k-1}c_a(m),
  \qquad \phi_a(0)=0,
\end{equation}
and the network potential
\begin{equation}
  \Phi(\bx)=\sum_{a\in\A}\phi_a(x_a)
  =\sum_{a\in\A}\sum_{m=0}^{x_a-1}c_a(m).
  \label{eq:discrete-potential}
\end{equation}
The term ``left endpoint'' refers to the fact that $\phi_a(k)$ evaluates $c_a$ at $0,1,\ldots,k-1$, the left endpoints of the $k$ unit intervals occupied by $k$ travelers. Rosenthal's potential for atomic congestion games \citep{Rosenthal1973IJGT,Rosenthal1973,MondererShapley1996} instead sums $c_a(1),\ldots,c_a(k)$, because a player there incurs the cost after joining the link. CSUE evaluates each candidate route before the traveler is inserted, so $\Phi$ is exactly Rosenthal's potential for the shifted link costs $\tilde c_a(k):=c_a(k-1)$ for $k\ge1$. The CSUE conditionals are therefore the logit (log-linear) revision rule of an atomic congestion game with costs $\tilde c_a$, and the Gibbs form of their joint law is the network instance of the result of \citet{Blume1993} for logit dynamics in potential games. The proof is included below.

\begin{proposition}[Joint Gibbs distribution of logit CSUE]
\label{thm:gibbs}
Under \cref{ass:core}, the probability mass function
\begin{equation}
  \pi_R(\bR)
  =\frac{1}{Z_R}
   \exp\{-\theta\Phi(\bx(\bR))\},
  \qquad \bR\in\mathcal S,
\end{equation}
with $Z_R=\sum_{\bR\in\mathcal S}\exp\{-\theta\Phi(\bx(\bR))\}\in(0,\infty)$, is the unique distribution on $\mathcal S$ with the full conditionals in \cref{eq:csue-conditional}.
\end{proposition}

\begin{proof}
The space $\mathcal S$ is finite and nonempty, and $\Phi(\bx(\bR))$ is finite for every $\bR\in\mathcal S$. Hence $0<Z_R<\infty$, and $\pi_R$ is positive on all of $\mathcal S$.

Fix all routes except $R_{wj}$. If traveler $(w,j)$ is inserted on route $r$, then every link $a\in r$ increases from $x_a^{-wj}$ to $x_a^{-wj}+1$. Therefore
\begin{align}
  \Phi(\bx^{-wj}+\Delta_r)-\Phi(\bx^{-wj})
  &=\sum_{a\in r}\left\{\phi_a(x_a^{-wj}+1)-\phi_a(x_a^{-wj})\right\}
  \notag\\
  &=\sum_{a\in r}c_a(x_a^{-wj})
  =C_{wr}(\bx^{-wj}).
\end{align}
As a function of $r$, $\pi_R(r,\bR_{-(wj)})$ is thus proportional to $\exp\{-\theta C_{wr}(\bx^{-wj})\}$, and normalizing over $\Rset_w$ gives \cref{eq:csue-conditional}.

For uniqueness, every conditional in Equation~\eqref{eq:csue-conditional} is strictly positive because all costs are finite, and $\mathcal S$ is a product space. Any distribution with these conditionals is therefore positive on all of $\mathcal S$, as can be seen by changing one traveler at a time from a profile of positive probability. Brook's lemma \citep{Brook1964,Besag1974} then writes $\pi(\bR)/\pi(\bR')$, for any two profiles, as a product of full-conditional ratios along a sequence of single-traveler changes from $\bR'$ to $\bR$. All such distributions hence share the same probability ratios, and normalization makes them equal to $\pi_R$.
\end{proof}

The partition function $Z_R$ may be expensive to evaluate, but sampling does not need it, because it cancels from every conditional. As $\theta\downarrow0$, the law tends to the uniform law on $\mathcal S$. As $\theta\to\infty$, it tends to the uniform law on the profiles that minimize $\Phi$; these are discrete potential minimizers, not in general a continuous-flow Wardrop equilibrium.

\begin{remark}[Logit dynamics and heterogeneous precision]
\label{rem:logit-dynamics}
If travelers revise at independent Poisson clocks with fixed positive rates and draw their new routes from Equation~\eqref{eq:csue-conditional}, the resulting process is reversible with stationary law $\pi_R$, since each revision is a single-traveler Gibbs update; this is the logit dynamics of \citet{Blume1993} for the congestion game with costs $\tilde c_a$. The common precision in \cref{ass:core}(ii) is needed for this construction: with traveler-specific precisions the conditionals are in general incompatible, as the two-traveler example in \cref{app:heterogeneous} shows.
\end{remark}%

\subsection{Connection with the Beckmann potential}
\label{subsec:beckmann}

For continuous flows, the Beckmann potential~\citep{Beckmann1956} is defined, whenever the link-cost integrals exist, by
\begin{equation}
  B(\bx)=\sum_{a\in\A}\int_0^{x_a}c_a(u)\,du.
\end{equation}
The energy $\Phi$ in \cref{eq:discrete-potential} is the left-endpoint Riemann sum of $B$ with unit step, not $B$ itself. A law proportional to $\exp\{-\theta B(\bx(\bR))\}$ would charge a traveler on link $a$ the averaged cost $\int_{x_a^{-wj}}^{x_a^{-wj}+1}c_a(u)\,du$, and the unshifted Rosenthal potential would charge $c_a(x_a^{-wj}+1)$, where CSUE charges $c_a(x_a^{-wj})$.

To obtain a continuous-flow limit, let each traveler represent a common physical flow quantum $\varepsilon>0$. In this scaling, $\bx$ denotes physical link flow: $k$ travelers contribute flow $k\varepsilon$ and incur link cost
$c_a(k\varepsilon)$. Thus $\bx(\bR)=\varepsilon\sum_{w,j}\Delta_{R_{wj}}$ and $\bx^{-wj}=\bx(\bR)-\varepsilon\Delta_{R_{wj}}$. We keep physical flow fixed while letting $\varepsilon$ tend to zero, so each traveler's contribution becomes negligible. For flows satisfying $x_a/\varepsilon\in\Nzero$ on every link, define
\begin{equation}
  \Phi_{\varepsilon}(\bx)
  =\sum_{a\in\A}\sum_{m=0}^{x_a/\varepsilon-1}c_a(m\varepsilon).
\end{equation}
Inserting one traveler on route $r$ gives $\Phi_\varepsilon(\bx^{-wj}+\varepsilon\Delta_r)-\Phi_\varepsilon(\bx^{-wj})=C_{wr}(\bx^{-wj})$. The proof of \cref{thm:gibbs} therefore applies to the scaled system with potential $\Phi_\varepsilon$ and flow increments $\varepsilon\Delta_r$. The scaled potential $\varepsilon\Phi_\varepsilon(\bx)$ is a left-endpoint Riemann sum for $B(\bx)$ with interval width $\varepsilon$. For continuous costs $\varepsilon\Phi_\varepsilon(\bx^\varepsilon)\to B(\bx)$ along any sequence of lattice flows $\bx^\varepsilon\to\bx$, and for nondecreasing costs $0\le B(\bx)-\varepsilon\Phi_\varepsilon(\bx)\le\varepsilon\sum_{a\in\A}[c_a(x_a)-c_a(0)]$ at every lattice flow (\cref{prop:beckmann-limit} in \cref{app:beckmann-limit}).

This limit concerns the potential only. At fixed $\theta$, the congestion exponent and the logarithm of the path-count multiplicity introduced in \cref{sec:pushforwards} both enter at the $1/\varepsilon$ scale, and their joint limit in \cref{subsec:large-population-sue} is deterministic logit SUE at fixed physical OD demand, not Wardrop equilibrium.

The quantum $\varepsilon$ is the physical size of one route decision, so $1/\varepsilon$ is the number of independent decision-makers per unit of demand. It sets the scale of all fluctuations: at fixed physical demand, the law concentrates on its limit as $\varepsilon$ decreases (\cref{thm:large-population-sue}), and physical-flow variances scale approximately linearly in $\varepsilon$ (\cref{rem:acceptance-scaling,subsec:exp-law}). A quantum larger than one vehicle represents travelers who choose routes in correlated groups, so $\varepsilon$ acts as a dispersion parameter that can be calibrated against observed count variability. Unless stated otherwise, the experiments of \cref{sec:experiments} use $\varepsilon=1$ in the demand units of each network; \cref{subsec:exp-law} varies it.

\section{Pushforward Flow Laws and the Deterministic SUE Limit}
\label{sec:pushforwards}

This section derives the laws of the aggregate flows: path counts and total link flows in \cref{subsec:pathflow-distribution}, their large-population limit in \cref{subsec:large-population-sue}, and, for acyclic bush route sets, the block-layered link flows on which the samplers of \cref{sec:sampling} operate (\cref{subsec:layered}).

\subsection{Path-count and total link-flow laws}
\label{subsec:pathflow-distribution}

Returning to the unit-traveler setting, for $w\in\W$ and
$r\in\Rset_w$, define the aggregate path count
\begin{equation}
    F_{wr}
    =
    \sum_{j=1}^{n_w}
    \mathbf{1}\{R_{wj}=r\}.
\end{equation}
Let $\bF=(F_{wr})$, and define the feasible integer path-flow set
\begin{equation}
    \Fset_n
    =
    \left\{
        \bfv\in\Nzero^{\sum_w|\Rset_w|}:
        \sum_{r\in\Rset_w}f_{wr}=n_w
        \text{ for all }w
    \right\}.
\end{equation}
A path-flow vector $\bfv\in\Fset_n$ induces the
link flow $\bx=\Delta\bfv$, where $\Delta$ is the incidence matrix of \cref{sec:model}. Let $\Xset=\{\Delta\bfv:\bfv\in\Fset_n\}$ be the set of feasible total link flows and, for $\bx\in\Xset$, let $\Dset(\bx):=\{\bfv\in\Fset_n:\Delta\bfv=\bx\}$ be the fiber of path counts that produce $\bx$. Sets of this kind are the latent state space of network tomography, where link counts are observed and route flows are not \citep{Hazelton2015}.

\begin{proposition}[Path-count and total link-flow laws]
\label{prop:pushforward}
Under \cref{ass:core}, the path counts and the total link flow induced by $\pi_R$ have the laws
\begin{align}
  \pi_F(\bfv)
  &=\frac{1}{Z_R}\exp\{-\theta\Phi(\Delta\bfv)\}\,M(\bfv),
  &M(\bfv)&:=\prod_{w\in\W}\frac{n_w!}{\prod_{r\in\Rset_w}f_{wr}!},
  &\bfv&\in\Fset_n,
  \label{eq:path-flow-law}\\
  \pi_X(\bx)
  &=\frac{1}{Z_R}\exp\{-\theta\Phi(\bx)\}\,\Gamma(\bx),
  &\Gamma(\bx)&:=\sum_{\bfv\in\Dset(\bx)}M(\bfv),
  &\bx&\in\Xset,
  \label{eq:link-flow-law}
\end{align}
where $M(\bfv)$ and $\Gamma(\bx)$ are the numbers of labeled route profiles that produce $\bfv$ and $\bx$.
\end{proposition}

\begin{proof}
All profiles with path counts $\bfv$ have link flow $\Delta\bfv$ and therefore the same probability under $\pi_R$, and the travelers of OD pair $w$ can be assigned to routes in $n_w!/\prod_{r\in\Rset_w}f_{wr}!$ ways. This gives $\pi_F$. Summing $\pi_F$ over the fiber $\Dset(\bx)$, on which $\Phi(\Delta\bfv)=\Phi(\bx)$, gives $\pi_X$. The fibers partition $\Fset_n$, so both laws keep the normalizing constant $Z_R$.
\end{proof}

The multiplicity $M(\bfv)$ arises exactly from aggregating labeled travelers into path counts. Dividing $M(\bfv)$ by $\prod_{w}|\Rset_w|^{n_w}$ gives the product-multinomial probability of $\bfv$ under uniform route choice, and $\pi_F$ is the exponential tilt of that law by $e^{-\theta\Phi}$. A similar decomposition underlies the sampler of \citet{Kleer2023}, which draws path counts and then assigns travelers to them uniformly. The mode of $\pi_F$ minimizes $\theta\Phi(\Delta\bfv)+\sum_{w\in\W}\sum_{r\in\Rset_w}\log(f_{wr}!)$ over $\Fset_n$ and is in general distinct both from the mean and from deterministic logit SUE.

Unlike $M$, the multiplicity $\Gamma(\bx)$ is not explicit. It sums over every path-flow decomposition of $\bx$, and for general finite route sets the sum can involve many allocations even for a single OD pair. \Cref{subsec:layered} identifies route-set structure under which the within-block count becomes explicit.

\subsection{Large-population connection to deterministic logit SUE}
\label{subsec:large-population-sue}

Following the flow-quantum scaling of \cref{subsec:beckmann}, fix physical OD demand $q_w>0$ and consider refinements $\varepsilon\downarrow0$ for which
\[
    n_w^\varepsilon=\frac{q_w}{\varepsilon}\in\mathbb{N}.
\]
If $N_{wr}^\varepsilon$ is the number of travelers using route $r$, define the corresponding physical path flow by $f_{wr}^\varepsilon=\varepsilon N_{wr}^\varepsilon$, so that $\sum_{r\in\Rset_w}f_{wr}^\varepsilon=q_w$. Let $\pi_F^\varepsilon$ denote the path-flow distribution in Equation~\eqref{eq:path-flow-law} of the scaled system, in which $\Phi$ is replaced by $\Phi_\varepsilon$ and $n_w$ by $n_w^\varepsilon$. Let $Z_F^\varepsilon$ denote its normalizing constant.

Define the continuous feasible set
\begin{equation}
    \mathcal F(\mathbf q)
    =
    \left\{
        \bfv\ge0:
        \sum_{r\in\Rset_w}f_{wr}=q_w
        \text{ for all }w
    \right\}.
\end{equation}
We use the convention $0\log0=0$.

\begin{proposition}[Large-population concentration and logit SUE]
\label{thm:large-population-sue}
Keep the finite route sets and $\theta>0$ fixed. Suppose that every link-cost function $c_a$ is continuous and nondecreasing. For any sequence of feasible lattice flows
$\bfv^\varepsilon\to\bfv\in\mathcal F(\mathbf q)$,
\begin{equation}
    -\varepsilon
    \log \pi_F^\varepsilon(\bfv^\varepsilon)
    =
    I(\bfv)+\kappa_\varepsilon+o(1),
    \label{eq:large-population-rate}
\end{equation}
where $\kappa_\varepsilon=\varepsilon\log Z_F^\varepsilon$ is independent of $\bfv$, and
\begin{equation}
    I(\bfv)
    =
    \theta B(\Delta\bfv)
    +
    \sum_{w\in\W}
    \sum_{r\in\Rset_w}
    f_{wr}\log\!\left(\frac{f_{wr}}{q_w}\right).
    \label{eq:large-population-rate-function}
\end{equation}
Writing $I_\star=\min_{\bfv\in\mathcal F(\mathbf q)}I(\bfv)$, we have $\kappa_\varepsilon\to-I_\star$, so the normalized limiting rate is $I(\bfv)-I_\star$. Consequently, $\pi_F^\varepsilon$ concentrates on the unique minimizer of
\begin{equation}
    \min_{\bfv\in\mathcal F(\mathbf q)}
    \left\{
        B(\Delta\bfv)
        +
        \frac{1}{\theta}
        \sum_{w\in\W}
        \sum_{r\in\Rset_w}
        f_{wr}\log\!\left(\frac{f_{wr}}{q_w}\right)
    \right\}.
    \label{eq:logit-sue-program}
\end{equation}
The minimizer $\bfv^\star$ satisfies
\begin{equation}
    f_{wr}^\star
    =
    q_w
    \frac{
        \exp\{-\theta C_{wr}(\bx^\star)\}
    }{
        \sum_{s\in\Rset_w}
        \exp\{-\theta C_{ws}(\bx^\star)\}
    },
    \qquad
    \bx^\star=\Delta\bfv^\star,
    \label{eq:logit-sue-fixed-point}
\end{equation}
which are the deterministic logit SUE fixed-point equations.
\end{proposition}

The proof, in \cref{app:large-population-details}, combines \cref{prop:beckmann-limit} for the congestion term with Stirling's formula for the multiplicity and a Laplace argument on the feasible lattice. Program~\eqref{eq:logit-sue-program} is the entropy-regularized Beckmann formulation of \citet{Fisk1980}, and the limit is consistent with \citet{Hazelton1998}. Continuity of $\bfv\mapsto\Delta\bfv$ gives concentration of physical total link flows at $\bx^\star$; boundedness of the feasible physical flows also gives convergence of path-flow and link-flow means to $\bfv^\star$ and $\bx^\star$. If $\theta\to\infty$ after this limit, every accumulation point of $\bfv^\star$ is a Wardrop equilibrium on the fixed route sets.

\subsection{Block-layered link flows on acyclic route sets}
\label{subsec:layered}

Retaining block labels gives explicitly evaluable unnormalized flow weights when each origin's route sets consist of all paths in a fixed acyclic subnetwork.

\begin{assumption}[Acyclic bush route sets]
\label{ass:bush}
For each origin $o$, fix an acyclic subnetwork $\mathcal T_o=(\V_o,\A_o)$ of $G$ containing every destination served from $o$, with every node reachable from $o$. For every OD pair $w$ with $o_w=o$, $\Rset_w$ is the set of all $o$--$d_w$ paths in $\mathcal T_o$.
\end{assumption}

\Cref{ass:bush} uses the acyclic route-set structure of Dial's logit loading \citep{Dial1971} and origin-based assignment \citep{BarGera2002}. 

Fix a partition of the travelers into blocks $g\in\mathcal G$ such that all travelers in block $g$ share an origin $o_g$. Origin blocks, which contain every traveler from one origin, and OD blocks, which contain every traveler of one OD pair, are two natural choices. Finer blocks that split an OD pair into labeled sub-batches are also allowed. For block $g$ and node $i$, let $n_i^g$ be the number of block-$g$ travelers whose destination is $i$, and define the block link-flow layer
\begin{equation}
  \bx^g(\bR)=\sum_{(w,j)\in g}\Delta_{R_{wj}},
  \qquad
  \bx(\bR)=\sum_{g\in\mathcal G}\bx^g(\bR).
\end{equation}
Write $\delta^-(i)$ and $\delta^+(i)$ for the incoming and outgoing links of node $i$ in $\mathcal T_{o_g}$. A layer is supported on $\A_{o_g}$ and satisfies flow conservation
\begin{equation}
  m_i^g(\bx^g)
  :=n_i^g+\sum_{a\in\delta^+(i)}x_a^g
  =\sum_{a\in\delta^-(i)}x_a^g,
  \qquad i\in\V_{o_g}\setminus\{o_g\}.
  \label{eq:layer-conservation}
\end{equation}
Let $\Xset_g$ be the set of nonnegative integer vectors on $\A_{o_g}$ satisfying Equation~\eqref{eq:layer-conservation}. Because $\mathcal T_{o_g}$ is acyclic, the flow decomposition theorem \citep{AhujaMagnantiOrlin1993} implies that every element of $\Xset_g$ is induced by some assignment of block-$g$ travelers to bush paths. Feasibility of a layer is checked node by node.

\begin{theorem}[Block-layered link-flow distribution]
\label{thm:layered}
Under \cref{ass:core,ass:bush}:
\begin{enumerate}[label=(\alph*)]
\item The number of labeled route assignments of block-$g$ travelers that produce the layer $\bx^g\in\Xset_g$ is
\begin{equation}
  \Omega_g(\bx^g)
  =\prod_{i\in\V_{o_g}\setminus\{o_g\}}
   \frac{m_i^g(\bx^g)!}{\prod_{a\in\delta^-(i)}x_a^g!}.
  \label{eq:layer-multiplicity}
\end{equation}
\item The joint distribution of the layers $(\bX^g)_{g\in\mathcal G}$ induced by $\pi_R$ is
\begin{equation}
  \pi_L\bigl((\bx^g)_{g}\bigr)
  =\frac{1}{Z_R}
   \exp\Bigl\{-\theta\,\Phi\Bigl(\sum_{g\in\mathcal G}\bx^g\Bigr)\Bigr\}
   \prod_{g\in\mathcal G}\Omega_g(\bx^g),
  \qquad \bx^g\in\Xset_g .
  \label{eq:layered-law}
\end{equation}
\item For each block $g$, with background flow $\bb^g:=\sum_{h\ne g}\bx^h$,
\begin{equation}
  \pi_L\bigl(\bx^g\given(\bx^h)_{h\ne g}\bigr)
  \propto
  \Omega_g(\bx^g)\,
  \exp\bigl\{-\theta\,[\Phi(\bb^g+\bx^g)-\Phi(\bb^g)]\bigr\},
  \qquad \bx^g\in\Xset_g.
  \label{eq:layer-conditional}
\end{equation}
\end{enumerate}
\end{theorem}

The proof is given in \cref{app:layered}. Part (a) counts assignments by choosing each traveler's incoming link at each node in reverse topological order. The single-OD finite factorial count appears in \citet[Section~1.2]{Akamatsu1997}; here it is stated for the labeled blocks above and used in the CSUE law.

The product formula gives unnormalized masses; the normalizing constant $Z_R$ remains unknown. Summing Equation~\eqref{eq:layered-law} over compatible layers recovers Equation~\eqref{eq:link-flow-law}: under \cref{ass:bush}
\[
  \Gamma(\bx)=\sum_{(\bx^g)_g:\ \sum_g\bx^g=\bx}\ \prod_{g\in\mathcal G}\Omega_g(\bx^g).
\]
Blocks sharing an origin can be combined into one origin layer by the same theorem; aggregating different origins generally requires a further decomposition sum. Paths are thus summed out within each block, and the only summation left in $\Gamma$ is over decompositions of $\bx$ into layers. The total-flow law does not depend on the block partition when the bushes are fixed.

For a two-destination example, take links $a=(o,v)$, $b=(o,u)$, $c=(u,v)$, and $e=(v,d)$, with one traveler destined for $v$ and one for $d$. Their route sets are $\{a,bc\}$ and $\{ae,bce\}$. With both travelers in one origin block, the feasible layers in link order $(a,b,c,e)$ are
\[
  \bx^g(k)=(k,2-k,2-k,1),\qquad
  \Omega_g(\bx^g(k))=\frac{2!}{k!(2-k)!},\qquad k=0,1,2.
\]
At $v$, two travelers must be assigned to the incoming links; the other non-origin nodes contribute a factor of one. For $k=1$, either destination's traveler may use $a$, so two distinct OD path-count vectors give the same layer. If $c_a(s)=2+s$ and $c_b(s)=c_c(s)=c_e(s)=1$, then $\Phi(\bx^g(k))=5+k(k-1)/2$, and the probabilities of these three layers are $(1,2,e^{-\theta})/(3+e^{-\theta})$. Since there is only one origin block, total link flow equals its layer and $\Gamma=\Omega_g$.

In the flow-quantum scaling of \cref{subsec:large-population-sue}, $-\varepsilon\log\Omega_g$ converges to the negative path-choice entropy of the backward Markov decomposition of \citet{Akamatsu1997}, and the layered law concentrates at the logit SUE layers (\cref{app:layer-limit}).

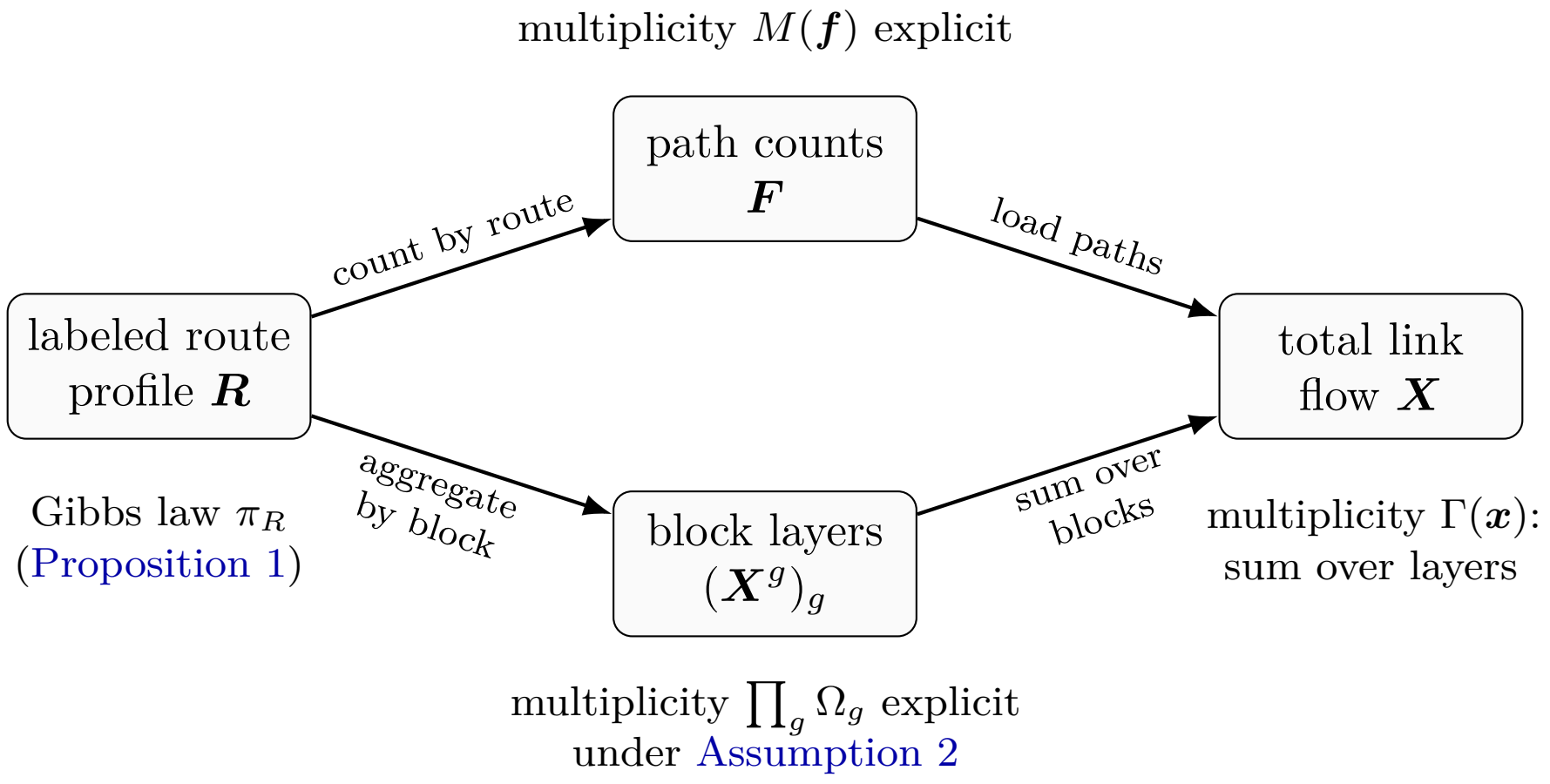
\begin{figure}[tbp]
\centering
\begin{tikzpicture}[
  node distance=19mm,
  box/.style={
      draw,
      rounded corners,
      align=center,
      minimum width=23mm,
      minimum height=11mm,
      fill=black!2,
      font=\small
  },
  arr/.style={-{Latex[length=2.2mm]},thick},
  lab/.style={font=\scriptsize,align=center,sloped,above,inner sep=2.5pt},
  note/.style={font=\footnotesize,align=center,text width=31mm}
]
  \node[box] (R) at (0,0) {labeled route\\profile $\bR$};
  \node[box] (F) at (4.6,1.5) {path counts\\$\bF$};
  \node[box] (L) at (4.6,-1.5) {block layers\\$(\bX^g)_{g}$};
  \node[box] (X) at (9.2,0) {total link\\flow $\bX$};

  \draw[arr] (R) -- node[lab]{count by route} (F);
  \draw[arr] (R) -- node[lab,below,pos=0.45]{aggregate\\by block} (L);
  \draw[arr] (F) -- node[lab]{load paths} (X);
  \draw[arr] (L) -- node[lab,below,pos=0.55]{sum over\\blocks} (X);

  \node[note,below=3mm of R] {Gibbs law $\pi_R$\\(\cref{thm:gibbs})};
  \node[note,above=2mm of F,text width=48mm] {multiplicity $M(\bfv)$ explicit};
  \node[note,below=2mm of L,text width=48mm] {multiplicity $\prod_g\Omega_g$ explicit\\under \cref{ass:bush}};
  \node[note,below=3mm of X] {multiplicity $\Gamma(\bx)$:\\sum over layers};
\end{tikzpicture}%

\caption{
Exact representations of finite-population logit CSUE. The maps are generally many-to-one. Path counts need not determine block layers when an OD pair is split into labeled sub-batches. Path-count multiplicities are explicit, as are layer multiplicities under \cref{ass:bush}; total-flow multiplicities sum over compatible layers. All four laws share $Z_R$. The samplers of \cref{sec:sampling} operate on the block layers.
}
\label{fig:pushforward-map}
\end{figure}

A Gibbs or Metropolis update on total link flow alone would have to evaluate ratios $\Gamma(\bx')/\Gamma(\bx)$ or draw a multicommodity decomposition given the total flow, and its proposals would have to preserve every OD demand. The layered state needs none of this: ratios of $\Omega_g$ are explicit, and feasibility reduces to the node-wise conservation Equation~\eqref{eq:layer-conservation}. The samplers of \cref{sec:sampling} take Equation~\eqref{eq:layered-law} as their target, update one block at a time from Equation~\eqref{eq:layer-conditional}, and report $\bx=\sum_g\bx^g$ as a derived quantity. The price is memory: one link-flow vector per block rather than one for the network, matching origin-based assignment when origin blocks are used. \Cref{fig:pushforward-map} summarizes the four representations.

\section{From Exact Gibbs Sampling to Path-Free Block Sampling}
\label{sec:sampling}

This section gives two samplers. The traveler-level Gibbs sampler of \cref{sec:exactgibbs} needs no bush structure but costs one update per traveler. The block sampler of \cref{subsec:block-conditional,subsec:frozen,subsec:mh,subsec:warmup} works on the layered state of \cref{thm:layered}: it proposes a whole block by path-free loading at frozen costs and corrects the proposal by an accept--reject step. Throughout, a sampler is called exact when its invariant distribution is the target law.

\subsection{Traveler-level Gibbs sampling}
\label{sec:exactgibbs}

If $\bR\sim\pi_R$, the induced link flow $\bX:=\bx(\bR)$ has law $\pi_X$. A Gibbs sweep on labeled route profiles visits the travelers in a fixed or state-independent random order. For traveler $(w,j)$ it removes the current route from the link flow, computes $C_{wr}(\bx^{-wj})$ for every $r\in\Rset_w$, draws a new route from \cref{eq:csue-conditional}, and adds it to the link flow. Each step draws one coordinate of $\bR$ from its full conditional under $\pi_R$ (\cref{thm:gibbs}), so the sweep leaves $\pi_R$ invariant. All full conditionals are positive on the product space $\mathcal S$, so one sweep moves any profile to any other with positive probability and retains the current profile with positive probability. The chain is therefore irreducible and aperiodic, and $\pi_R$ is its unique invariant distribution.

This sampler applies to the fixed finite route sets of \cref{sec:model}, but a sweep costs $N$ route-choice updates, each requiring the costs of all candidate routes, and the state stores $N$ routes. Under \cref{ass:bush}, each update can be carried out without enumerating routes by the recursion of \cref{subsec:frozen} applied to a single traveler. The $N$ updates per sweep and the labeled state remain.

\subsection{Block updates and the block conditional}
\label{subsec:block-conditional}

To reduce the cost of a sweep, we update all travelers of a block jointly and represent the block only by its link-flow layer. Throughout the rest of this section, \cref{ass:bush} holds, the bushes $\mathcal T_o$ and the block partition $\mathcal G$ of \cref{subsec:layered} are fixed, and the state of the sampler is the collection of layers $(\bx^g)_{g\in\mathcal G}$. For block $g$, write $\mathcal T_g=(\V_g,\A_g)$ for the bush of its origin $o_g$, $\delta^-(i)$ and $\delta^+(i)$ for the incoming and outgoing links of node $i$ in $\mathcal T_g$, and
\begin{equation}
  \bb^g=\bx-\bx^g=\sum_{h\ne g}\bx^h
\end{equation}
for the leave-block-out background flow. By \cref{thm:layered}(c), the conditional distribution of a candidate layer $\bz\in\Xset_g$ given the other layers is
\begin{equation}
  \pi_g(\bz\given\bb^g)
  \propto
  \Omega_g(\bz)\,
  \exp\bigl\{-\theta\,[\Phi(\bb^g+\bz)-\Phi(\bb^g)]\bigr\}.
  \label{eq:exact-block}
\end{equation}
Aggregation introduces no approximation: the layers outside the block enter only through $\bb^g$, and $\Omega_g(\bz)$ counts the labeled route assignments that produce $\bz$.

By \cref{eq:discrete-potential}, the congestion increment in \cref{eq:exact-block} is
\begin{equation}
    \Phi(\bb^g+\bz)
    -
    \Phi(\bb^g)
    =
    \sum_{a\in\A}
    \sum_{k=0}^{z_a-1}
    c_a(b_a^g+k),
\end{equation}
with an empty inner sum interpreted as zero. Successive block travelers on link $a$ are charged the costs at flows $b_a^g,b_a^g+1,\ldots,b_a^g+z_a-1$, so the travelers of a block are coupled through the congestion they create among themselves, and Equation~\eqref{eq:exact-block} generally does not factor into independent route choices under a common cost field. The feasible layers are too many to enumerate. \Cref{subsec:frozen} constructs a path-free proposal from frozen costs, and \cref{subsec:mh} corrects it by an accept--reject step.

\subsection{Frozen-cost loading as a path-free proposal}
\label{subsec:frozen}

Fix a reference flow $\by\in\mathbb R_{\ge0}^{|\A|}$ and freeze all link costs at $\mathbf c(\by)$, evaluating $c_a$ at real arguments as in the Bureau of Public Roads (BPR) cost function. If the travelers of block $g$ chose bush routes independently with logit probabilities at these frozen costs, their layer would follow a law that we call the frozen-cost surrogate and denote by $\widetilde\pi_g(\cdot\given\by)$. This law can be described and sampled without enumerating routes.

Let $\omega_a(\by)=\exp\{-\theta c_a(y_a)\}$ for $a\in\A_g$, and define node weights by the forward topological recursion
\begin{equation}
  H_{o_g}(\by)=1,\qquad
  H_i(\by)=\sum_{a=(j,i)\in\delta^-(i)}H_j(\by)\,\omega_a(\by),
  \quad i\ne o_g .
  \label{eq:block-forward-partition}
\end{equation}
$H_i(\by)$ is the total frozen-cost weight of all bush paths from $o_g$ to $i$, so $H_d(\by)$ is the logit denominator of OD pair $(o_g,d)$. The predecessor probabilities
\begin{equation}
  \gamma_{(j,i)}(\by)=\frac{H_j(\by)\,\omega_{(j,i)}(\by)}{H_i(\by)},
  \qquad (j,i)\in\delta^-(i),
  \label{eq:predecessor-prob}
\end{equation}
sum to one over $\delta^-(i)$. For any bush path $r$ from $o_g$ to $d$, their product telescopes to $\prod_{a\in r}\gamma_a(\by)=\exp\{-\theta C_{wr}(\by)\}/H_d(\by)$, the frozen-cost logit probability of $r$.

\begin{proposition}[Path-free sampling of the frozen-cost surrogate]
\label{prop:surrogate}
Under \cref{ass:core,ass:bush}, for any reference flow $\by$:
\begin{enumerate}[label=(\alph*)]
\item the layer law of independent frozen-cost logit choices is
\begin{equation}
  \widetilde\pi_g(\bz\given\by)
  =\frac{\Omega_g(\bz)\,\exp\{-\theta\,\mathbf c(\by)^{\trans}\bz\}}
        {\prod_{i\in\V_g}H_i(\by)^{n_i^g}},
  \qquad \bz\in\Xset_g ;
  \label{eq:surrogate-law}
\end{equation}
\item processing the nodes $i\ne o_g$ in reverse topological order and drawing
\begin{equation}
  (Z_a)_{a\in\delta^-(i)}
  \sim
  \operatorname{Multinomial}\Bigl(m_i^g(\bZ),\,(\gamma_a(\by))_{a\in\delta^-(i)}\Bigr),
  \qquad
  m_i^g(\bZ)=n_i^g+\sum_{a\in\delta^+(i)}Z_a ,
  \label{eq:block-reverse-loading}
\end{equation}
produces $\bZ\sim\widetilde\pi_g(\cdot\given\by)$.
\end{enumerate}
The recursion Equation~\eqref{eq:block-forward-partition} and the loading Equation~\eqref{eq:block-reverse-loading} each visit every node and link of $\mathcal T_g$ a bounded number of times.
\end{proposition}

The proof is given in \cref{app:surrogate}. Part (b) is an integer, sampling counterpart of the loading of \citet{Dial1971}, run from the destinations toward the origin. When node $i$ is processed, the counts on its outgoing links are already known, so $m_i^g$ combines travelers ending at $i$ with travelers passing through it. The predecessor probabilities do not depend on a traveler's destination, so travelers of different destinations that meet at a node are split by a single multinomial draw. An OD block is the special case of one destination; an origin block uses the same recursion with several. Replacing the draws in Equation~\eqref{eq:block-reverse-loading} by their expectations, $Z_a=m_i^g\gamma_a(\by)$, gives the expected layer $\E_{\widetilde\pi_g(\cdot\mid\by)}[\bZ]$ at the same cost.

Comparing Equation~\eqref{eq:surrogate-law} with Equation~\eqref{eq:exact-block} shows that the block conditional is an exponential tilt of the surrogate. With the residual
\begin{equation}
  \rho_g(\bz;\bb,\by)
  :=\Phi(\bb+\bz)-\Phi(\bb)-\mathbf c(\by)^{\trans}\bz
  =\sum_{a\in\A}\sum_{k=0}^{z_a-1}\bigl[c_a(b_a+k)-c_a(y_a)\bigr],
  \label{eq:rho}
\end{equation}
we have
\begin{equation}
  \pi_g(\bz\given\bb)
  =\frac{\widetilde\pi_g(\bz\given\by)\,\exp\{-\theta\rho_g(\bz;\bb,\by)\}}
        {\E_{\widetilde\pi_g(\cdot\mid\by)}\bigl[\exp\{-\theta\rho_g(\bZ;\bb,\by)\}\bigr]} .
  \label{eq:tilt}
\end{equation}
The multiplicity $\Omega_g$ and the node weights cancel, so the tilt depends on the layer only through the residual $\rho_g$: the surrogate and the block conditional differ only because costs are frozen. The residual vanishes identically for flow-independent costs and for a single-traveler block with $\by=\bb$; in the latter case the surrogate is the CSUE conditional Equation~\eqref{eq:csue-conditional}.

\subsection{Correction by Metropolis--Hastings}
\label{subsec:mh}

Because the residual is explicit, the discrepancy in Equation~\eqref{eq:tilt} can be removed by an accept--reject step. Let the reference flow be a deterministic function of the background only, $\by=\by(\bb^g)$. The background is unchanged by the move, so the reverse proposal uses the same reference. Given the current layer $\bx^g$, draw a proposal $\bz'\sim\widetilde\pi_g(\cdot\given\by)$ by \cref{prop:surrogate}(b) and accept it with probability
\begin{equation}
  \alpha_g(\bx^g,\bz')
  =\min\Bigl\{1,\ \exp\bigl(-\theta\,\bigl[\Phi(\bb^g+\bz')-\Phi(\bb^g+\bx^g)-\mathbf c(\by)^{\trans}(\bz'-\bx^g)\bigr]\bigr)\Bigr\}.
  \label{eq:mh-acceptance}
\end{equation}
This is the independence Metropolis--Hastings ratio \citep{Hastings1970} for target Equation~\eqref{eq:exact-block} and proposal Equation~\eqref{eq:surrogate-law}. The multiplicities and node weights cancel, so the ratio requires only the potential difference between two layers and a linear term, both evaluated on $\A_g$.

In the two-destination example of \cref{subsec:layered}, take $\bb=\by=\mathbf0$. Frozen loading draws $k\sim\operatorname{Binomial}(2,1/2)$, giving proposal probabilities $(1,2,1)/4$, whereas the target probabilities are $(1,2,e^{-\theta})/(3+e^{-\theta})$. For the move $k=1\to2$, the target ratio is $e^{-\theta}/2$ and the reverse-to-forward proposal ratio is $2$, so the acceptance probability is $e^{-\theta}$. At $\theta=\log2$, it is $1/2$.

\begin{theorem}[Exactness of the block sampler]
\label{prop:block-mh}
Assume \cref{ass:core,ass:bush}, fixed bushes, a fixed block partition, and a reference rule $\by(\cdot)$ that depends on the state only through $\bb^g$. Then each block update leaves the layered law $\pi_L$ in \cref{eq:layered-law} invariant. A sweep over the blocks in a fixed or state-independent random order is irreducible and aperiodic on $\prod_g\Xset_g$, so $\pi_L$ is its unique invariant distribution, and $\bx=\sum_g\bx^g$ has stationary distribution $\pi_X$.
\end{theorem}

The proof is given in \cref{app:block-mh}. If the accept--reject step is omitted, every proposal is accepted; we call this the \emph{unadjusted} frozen-cost sampler. It costs the same apart from evaluating Equation~\eqref{eq:mh-acceptance}, but its stationary law need not equal $\pi_L$. With a single block and a fixed reference flow, the unadjusted sampler returns independent draws from $\widetilde\pi_g(\cdot\given\by)$, that is, independent logit route choice at the frozen costs $\mathbf c(\by)$.

\begin{proposition}[Rejection rate and proposal error]
\label{prop:block-error-bound}
Under \cref{ass:core,ass:bush}, fix a block $g$, a feasible background $\bb$, and a reference flow $\by$. Suppose the current layer has law $\pi_g(\cdot\given\bb)$ and the proposal is drawn independently from $\widetilde\pi_g(\cdot\given\by)$, and let $\bar\alpha_g$ be the mean acceptance probability. Then
\begin{equation}
  \bigl\|\pi_g(\cdot\given\bb)-\widetilde\pi_g(\cdot\given\by)\bigr\|_{\mathrm{TV}}
  \le 1-\bar\alpha_g
  \le 2\,\bigl\|\pi_g(\cdot\given\bb)-\widetilde\pi_g(\cdot\given\by)\bigr\|_{\mathrm{TV}}.
\end{equation}
\end{proposition}

The proof, in \cref{app:block-error-bound}, uses only that the proposal does not depend on the current layer. The acceptance rate is therefore an observable diagnostic: at stationarity, the mean rejection rate of a block lies between one and two times the total variation distance between its proposal and its block conditional, which is the error that the unadjusted update makes in that conditional.\footnote{Accepted self-proposals count as acceptances.} Averaged over backgrounds at joint stationarity, the bound controls the average conditional discrepancy, but not full-sweep mixing or the invariant-law error of the unadjusted chain.

\paragraph{Choice of the reference flow.}
Every reference rule that depends on the state only through $\bb^g$ yields an exact sampler; the rule affects efficiency only. The leave-block-out choice $\by=\bb^g$ mirrors the leave-one-out convention of CSUE and makes the proposal equal to the block conditional for single-traveler blocks, for which every proposal is accepted and the update reduces to a traveler-level Gibbs step (\cref{sec:exactgibbs}). For large blocks, however, it ignores the congestion that the block imposes on itself. If $0\le c_a(u+1)-c_a(u)\le L_a$, Equation~\eqref{eq:rho} gives $0\le\rho_g(\bz;\bb^g,\bb^g)\le\frac12\sum_{a\in\A}L_az_a(z_a-1)$, so the residual grows quadratically with the number of block travelers on a link. The proposal then loads the block as if its own flow were absent, which can produce a substantial mismatch in means. In the extreme case of a network with a single origin and an origin block, $\bb^g=\mathbf 0$ and the proposal is free-flow logit loading, whatever the level of congestion.

For blocks containing more than one traveler, we construct a mean-field reference $\by=\bb^g+\hat\bz(\bb^g)$ using expected loadings. The equation $\bz=\E_{\widetilde\pi_g(\cdot\mid\bb^g+\bz)}[\bZ]$ is a deterministic block-SUE condition against $\bb^g$. Under continuous costs it has a solution, because expected loading is a continuous self-map of the compact convex set $\operatorname{conv}(\Xset_g)$. Choose a fixed $K\in\Nzero$ and predetermined damping values $0\le\eta_k\le1$, and run
\[
  \hat\bz^{(k+1)}=(1-\eta_k)\,\hat\bz^{(k)}+\eta_k\,\E_{\widetilde\pi_g(\cdot\mid\bb^g+\hat\bz^{(k)})}[\bZ],
  \qquad
  \hat\bz^{(0)}=\E_{\widetilde\pi_g(\cdot\mid\bb^g)}[\bZ],
\]
for $k=0,\ldots,K-1$, without warm starts from earlier block updates, and set $\hat\bz=\hat\bz^{(K)}$. The iterates remain in $\operatorname{conv}(\Xset_g)$ and depend only on $\bb^g$. Any finite $K$ gives a valid proposal, but centering accuracy affects acceptance and mixing.

\begin{remark}[Heuristic scaling of the acceptance rate]
\label{rem:acceptance-scaling}
Suppose costs are continuously differentiable and nondecreasing, and assume a common-mean local Gaussian approximation about $\hat\bz$, with $\by=\bb+\hat\bz$. In count coordinates, let $U_g$ have orthonormal columns spanning differences between feasible layers, and write
\[
  \Sigma_g=U_g^{\trans}\operatorname{Cov}_{\widetilde\pi_g}(\bZ)U_g,
  \qquad
  D_g=U_g^{\trans}\diag\{c_a'(y_a)\}U_g.
\]
The smooth approximation to the discrete congestion potential adds $\theta D_g$ to the proposal precision, giving approximate target covariance $(\Sigma_g^{-1}+\theta D_g)^{-1}$. Since $D_g\succeq0$ for nondecreasing costs, this target covariance is no larger than $\Sigma_g$: the surrogate omits the congestion feedback of the block on itself and is overdispersed. The eigenvalues $\lambda_k$ of $\theta\Sigma_g^{1/2}D_g\Sigma_g^{1/2}$ describe the curvature mismatch, including link-flow correlations. Small $\sum_k\lambda_k^2$ suggests high acceptance within this centered Gaussian approximation.

Under the scaling of \cref{subsec:large-population-sue}, count covariance is $O(1/\varepsilon)$ and count-scale cost derivatives are $O(\varepsilon)$, so the curvature eigenvalues can remain $O(1)$ as $\varepsilon\downarrow0$. Centering is harder: physical fluctuations are $O(\sqrt\varepsilon)$, a fixed number $K$ of mean-field iterations does not control the proposal mean on that scale, and acceptance can then fall as the population grows, which is why large blocks are split in practice.
\end{remark}%

\subsection{Warm-up, complete sampler, and complexity}
\label{subsec:warmup}

\Cref{prop:block-mh} conditions on fixed bushes and a fixed block partition. We determine both during a warm-up phase whose draws are discarded. For the bushes, any rule that produces, for each origin, an acyclic subnetwork in which every retained node, including every demanded destination, is reachable from the origin satisfies \cref{ass:bush}. An example is Dial's efficient-link rule: link $(i,j)$ belongs to $\mathcal T_o$ if $r_o(i)<r_o(j)$, where $r_o$ denotes shortest-path distances from $o$. For this construction, use strictly positive reference link costs, omit unreachable nodes, and require every demanded destination to be reachable. The reference costs may be updated from warm-up draws, or taken from a deterministic user-equilibrium or logit SUE solution computed beforehand; the latter makes the route sets independent of the sampler's randomness. For the partition, each origin starts as one block, and blocks whose acceptance rate falls below a target are split into OD blocks or labeled sub-batches. At the end of warm-up, the bushes and the partition are frozen, and each layer is reinitialized by one draw from its surrogate so that it lies in $\Xset_g$. We recompute the total flow and discard a further equilibration period under the frozen kernel before retaining draws. Chains used to assess the same total-flow law share the frozen bushes.

\begin{algorithm}[tbp]

\caption{Path-free block Metropolis--Hastings update}
\label{alg:aggregate-bush-sampler}
\begin{algorithmic}[1]
\Require Block $g$ with frozen bush $\mathcal T_g$, topological order, and demands $(n_i^g)_i$; current layer $\bx^g$ and consistent total flow $\bx$
\Require Precision $\theta$, cost and potential evaluators; fixed $K\in\Nzero$ and damping values $(\eta_k)_{k=0}^{K-1}$
\Ensure Updated layer and total flow, and acceptance indicator $\mathsf{acc}$
\State $b_a^g\gets x_a-x_a^g$ for $a\in\A_g$
\If{$\sum_{i\in\V_g}n_i^g=1$}
    \State $\by\gets\bb^g$
\Else
    \State Compute $\hat\bz(\bb^g)$ by the fixed initialization and $K$ damped iterations above
    \State $\by\gets\bb^g+\hat\bz(\bb^g)$
\EndIf
\State Compute $H(\by)$ and $\gamma(\by)$ in log space by \cref{eq:block-forward-partition,eq:predecessor-prob}
\State Set $\bz'\gets\mathbf 0$ on $\A_g$
\For{nodes $i\ne o_g$ of $\mathcal T_g$ in reverse topological order}
    \State $m_i\gets n_i^g+\sum_{a\in\delta^+(i)}z'_a$
    \State Draw $(z'_a)_{a\in\delta^-(i)}\sim\operatorname{Multinomial}\bigl(m_i,(\gamma_a(\by))_{a\in\delta^-(i)}\bigr)$
\EndFor
\State Set $\log\alpha_g(\bx^g,\bz')=0$ for a singleton block; otherwise evaluate it directly from \cref{eq:mh-acceptance}
\State Draw $U\sim\mathrm{Uniform}(0,1)$ and set $\mathsf{acc}\gets0$
\If{$\log U\le\log\alpha_g(\bx^g,\bz')$}
    \State $x_a\gets b_a^g+z'_a$ for $a\in\A_g$, leaving other entries unchanged
    \State $\bx^g\gets\bz'$, \quad $\mathsf{acc}\gets1$
\EndIf
\State \Return updated $(\bx^g,\bx,\mathsf{acc})$
\Statex \textit{Cost:} $O((K+1)(|\V_g|+|\A_g|))$ plus variate generation and potential evaluation.
\end{algorithmic}
\end{algorithm}

A sweep applies \cref{alg:aggregate-bush-sampler} once to every block, in a fixed or state-independent random order, and records $\bx$ and any functionals of interest. OD and origin sampling differ only in the block partition and the terminating-demand vectors; the loading mechanism is identical. All link-flow vector operations in the algorithm are restricted to $\A_g$. If current-flow costs are cached, an accepted move requires refreshing them only on changed links; the proposal costs are evaluated separately at $\by$.

\paragraph{Complexity.}
For a block on bush $\mathcal T_g$, the forward recursion, reverse loading, and each expected loading visit $O(|\V_g|+|\A_g|)$ nodes and links. A sweep costs
\[
  O\!\left((K+1)\sum_{g\in\mathcal G}(|\V_g|+|\A_g|)\right)
\]
arithmetic operations, plus random-variate generation and potential-difference evaluation; the latter costs $O(|\A_g|)$ with closed-form link potentials, as for BPR costs with an integer exponent, and $O(\sum_{a\in\A_g}|z'_a-x_a^g|)$ by direct summation otherwise. There is one block per active origin under origin blocking, one per OD pair under OD blocking, and $N$ under singleton blocking, so splitting blocks moves the cost toward the population size. Layer storage is $O(\sum_g|\A_g|)$. Traveler-level Gibbs sampling instead requires $N$ route-choice updates per sweep and stores $N$ routes, whether or not each update uses path-free loading.

\section{Numerical Experiments}
\label{sec:experiments}

The experiments address four questions: whether the block sampler reproduces the CSUE law (\cref{subsec:exp-validation}), what it costs on networks of realistic size (\cref{subsec:exp-performance}), how the CSUE law differs from deterministic SUE and from independent route choice (\cref{subsec:exp-law}), and what its covariance can be used for (\cref{subsec:exp-sensors}).

\subsection{Networks and settings}

\Cref{tab:exp-networks} summarizes the four networks used in the experiments.
The two-destination Braess network is small enough for the CSUE law to be
computed by exact enumeration. The Nguyen--Dupuis network
\citep{NguyenDupuis1984} provides a small test case on which long
traveler-level Gibbs runs remain inexpensive. Sioux Falls
\citep{LeBlancMorlokPierskalla1975} and Chicago Sketch are taken from
the Transportation Networks for Research repository \citep{TNTP}.

\begin{table}[htbp]
\centering
\caption{Test networks. $N$ is the number of travelers at the default
flow quantum $\varepsilon=1$; $\overline{|\A_o|}$ is the mean number of
links in an origin bush.}
\label{tab:exp-networks}
\small
\begin{tabular}{@{}lrrrrrr@{}}
\toprule
Network & $|\V|$ & $|\A|$ & Origins & OD pairs & $N$
& $\overline{|\A_o|}$ \\
\midrule
Braess, two destinations
& 4 & 5 & 1 & 2 & 60 & 5.00 \\
Nguyen--Dupuis
& 13 & 19 & 2 & 4 & 2{,}000 & 16.00 \\
Sioux Falls
& 24 & 76 & 24 & 528 & 360{,}600 & 37.42 \\
Chicago Sketch
& 933 & 2{,}950 & 386 & 52{,}322 & 1{,}137{,}496 & 786.70 \\
\bottomrule
\end{tabular}
\end{table}

All four networks use the BPR link cost
\(c_a(x)=t_a^0[1+0.15(x/u_a)^4]\). For the Braess network, the
free-flow times are \(\boldsymbol t^0=(1,1.3,0.4,1.2,0.9)\), and the
capacities are \(\boldsymbol u=(20,20,10,20,20)\). The corresponding
link order is \((o,a),(o,b),(a,b),(a,d),(b,d)\). Sioux Falls and Chicago Sketch use the
corresponding parameters in the Transportation Networks for Research
data files \citep{TNTP}.
Travel times are measured in minutes, so \(\theta\) is measured in
inverse minutes.

Bushes are constructed once by the strict efficient-link rule of \cref{subsec:warmup}, using deterministic-UE link
costs, and are then shared by all samplers on the same network. Unless
varied explicitly, the block proposal uses \(K=3\) mean-field iterations
with damping \(\eta_k=1/(k+2)\). For the large-network production runs,
a 100-sweep calibration stage splits a block if its pooled acceptance
rate is below 0.2. 

Each MCMC configuration uses four single-threaded chains with independent random-number streams. Two chains are initialized by legal integer loadings at SUE costs and two by legal integer loadings at free-flow costs. Demand integerization, software, and hardware are described in \cref{app:implementation}.
\FloatBarrier

\subsection{Validation against reference laws}
\label{subsec:exp-validation}

\paragraph{Braess network.}
The Braess network has nodes $o,a,b,d$ and links $(o,a)$, $(o,b)$,
$(a,b)$, $(a,d)$, and $(b,d)$. It is acyclic and serves as its own
bush. OD pair $(o,d)$ has three routes and $n_{od}=40$ travelers; OD
pair $(o,b)$ has two routes and $n_{ob}=20$ travelers. The path-count
space has $\binom{42}{2}\times21=18{,}081$ states, so $\pi_F$, $\pi_L$,
and $\pi_X$ are computed exactly from Equation~\eqref{eq:path-flow-law}
by enumeration. We run the traveler-level Gibbs sampler, the block MH
sampler with OD blocks and with the origin block, and the unadjusted
sampler with the origin block, each using four chains with $10^6$ retained sweeps per chain after 10,000 burn-in sweeps. With a single origin block the
background flow is zero and the reference flow is fixed, so the
unadjusted sampler returns independent draws from the frozen-cost
surrogate: it is independent logit route choice at the mean-field costs
(\cref{subsec:mh}). For each sampler, \cref{tab:exp-braess} reports the
total variation distance between the empirical and exact laws of $\bX$,
together with the maximum relative errors of link-flow means and
standard deviations and the maximum absolute error of link-flow
correlations. An i.i.d.\ sample from the exact law, matched in effective
sample size, provides a Monte Carlo reference. Deterministic logit SUE
flows are also compared with the exact means.

\begin{table}[tbp]
\centering
\caption{Errors against the exact CSUE law on the Braess network
($\theta=0.3$). TV is the total variation distance between the empirical
and exact laws of $\bX$; Mean and Std.\ dev.\ are maximum relative errors
over links; Corr.\ is the maximum absolute error over link pairs. These
four columns are in units of $10^{-3}$. Acc.\ is the mean acceptance rate
of block proposals.}
\label{tab:exp-braess}
\small
\begingroup
\setlength{\tabcolsep}{4pt}
\begin{tabular}{@{}lccccc@{}}
\toprule
 & TV & Mean & Std.\ dev. & Corr. & Acc. \\
\midrule
Traveler-level Gibbs & 2.24 & 0.10 & 0.69 & 0.43 & -- \\
Block MH, OD blocks & 2.65 & 0.19 & 0.49 & 0.51 & 0.819 \\
Block MH, origin block & 3.00 & 0.09 & 1.01 & 0.35 & 0.703 \\
Independent choice at mean-field costs & 221.8 & 5.87 & 483.7 & 155.0 & -- \\
Deterministic logit SUE & -- & 0.65 & -- & -- & -- \\
\midrule
i.i.d.\ sample from $\pi_X$ & 3.00 & 0.14 & 0.82 & 0.66 & -- \\
\bottomrule
\end{tabular}
\endgroup
\end{table}

The corrected block samplers reproduce the exact law up to Monte
Carlo error: their errors are of the same size as those of the i.i.d.\
sample. Independent choice at mean-field costs matches the mean flows to
within 0.6\%, but its total variation distance from the CSUE law is
0.222, it overstates link-flow standard deviations by up to 48\%, and
its correlations are off by up to 0.155. Deterministic logit SUE is
within 0.1\% of the CSUE mean and provides no variances or correlations.

The table also gives a direct check of
\cref{prop:block-error-bound}. For the origin block, the proposal is the
law sampled in the fourth row and the block conditional is the
CSUE law itself, so the total variation distance between them is the
0.222 of that row. The proposition then places the stationary rejection
rate of the origin-block sampler between 0.222 and 0.444; the observed
rate is 0.297.

\begin{figure}[tbp]
\centering
\captionsetup{font={small,stretch=1.0}}
\includegraphics[width=\linewidth]{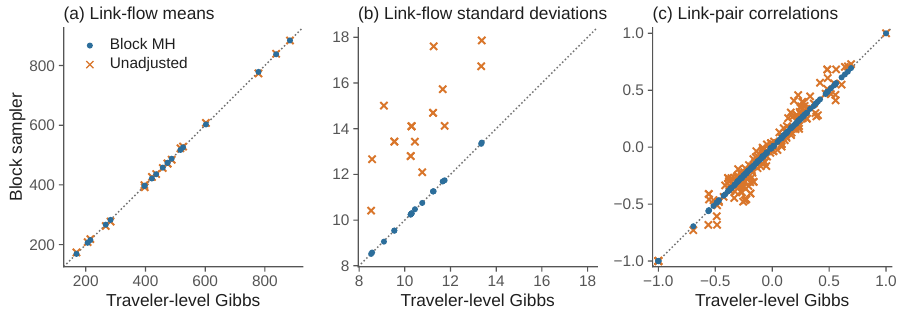}
\caption{Validation on the Nguyen--Dupuis network ($N=2{,}000$,
$\theta=0.3$). Origin-block MH (circles) and its unadjusted variant
(crosses) are compared with traveler-level Gibbs for (a) link-flow
means, (b) link-flow standard deviations, and (c) pairwise link-flow
correlations. Each point represents one link in (a)--(b) or one link
pair in (c). Dotted lines indicate equality.}
\label{fig:exp-validation}
\end{figure}

\paragraph{Nguyen--Dupuis network.}
When enumeration is infeasible, long runs of the traveler-level Gibbs
sampler provide the reference. On the Nguyen--Dupuis network with
$N=2{,}000$ travelers, we compare the origin-block MH sampler and its
unadjusted variant with traveler-level Gibbs, using four chains with $10^5$ retained sweeps per chain after 10,000 burn-in sweeps.
\Cref{fig:exp-validation}(a)--(c) compares link-flow means, standard
deviations, and pairwise correlations, respectively.

The corrected block sampler closely follows the identity line for all
three statistics in \cref{fig:exp-validation}. Its maximum relative
differences from the Gibbs reference are 0.0294\% for means and 0.346\%
for standard deviations; the largest absolute correlation difference
is 0.00560. The unadjusted sampler has mean-flow differences below
3.0\%, but overestimates the standard deviation on every link by
12.4\%--65.2\% and changes pairwise correlations by up to 0.229.
The direction of the error is consistent with the local
Gaussian comparison in \cref{rem:acceptance-scaling}: around a
well-centered block mean, the frozen-cost proposal has covariance
$\Sigma_g$, whereas the block conditional has approximate
covariance $(\Sigma_g^{-1}+\theta D_g)^{-1}$.
\FloatBarrier

\subsection{Sampler performance}
\label{subsec:exp-performance}

\Cref{tab:exp-performance} reports four block-sampler
configurations run at the common precision $\theta=0.3$. For each
configuration, it gives the number of blocks in the frozen production
kernel, the mean acceptance rate over blocks and retained sweeps, the
smallest bulk effective sample size over nonconstant link flows per
1,000 pooled retained draws, the largest rank-normalized
$\widehat R$ over those flows \citep{Vehtari2021}, and the time per
executed sweep. The last column gives the time per sweep of
traveler-level Gibbs sampling on the same bushes where that benchmark
was run.

Balanced-origin blocking on Chicago Sketch divides the travelers of each origin into labeled sub-blocks of at most 500 travelers. Within an origin, sub-block sizes differ by at most one traveler, and every OD demand is preserved. The Sioux Falls block sampler and the Chicago Sketch balanced-origin sampler each retain 500 sweeps per chain after 100 burn-in sweeps. The Sioux Falls traveler-level Gibbs reference retains 350 sweeps per chain after 50 burn-in sweeps. Effective sampling rates divide bulk ESS by total run time across chains, including burn-in: elapsed time for Nguyen--Dupuis and CPU time for Sioux Falls.

\begin{table}[!htbp]
\centering
\caption{Performance of the block-sampler configurations at
$\theta=0.3$. Acceptance is averaged over blocks and retained sweeps.
ESS/1,000 is the minimum link-flow bulk ESS divided by the total number
of retained draws across four chains and multiplied by 1,000.
$\widehat R$ is the maximum over nonconstant link flows. Times include
discarded production sweeps. The two Chicago Sketch rows use balanced
origin sub-blocks and OD blocks, respectively. OD blocking includes labeled sub-blocks, so the block count can exceed the number of OD pairs.}
\label{tab:exp-performance}
\small
\begingroup
\setlength{\tabcolsep}{3.5pt}
\begin{tabular}{@{}llrrrrrr@{}}
\toprule
Network & Blocking & Blocks & Acceptance & ESS/1,000 & $\max\widehat R$ &
\shortstack{Block\\s/sweep} & \shortstack{Gibbs\\s/sweep} \\
\midrule
Nguyen--Dupuis
& origin
& 2
& 0.470
& 225.4
& 1.0001
& $7.65\times10^{-5}$
& $4.34\times10^{-3}$ \\

Sioux Falls
& OD
& 668
& 0.948
& 635.5
& 1.0031
& $5.61\times10^{-3}$
& 0.706 \\

Chicago Sketch
& origin
& 2{,}469
& 0.982
& 670.1
& 1.0062
& 0.883
& -- \\

Chicago Sketch
& OD
& 57{,}494
& 0.9987
& 713.1
& 1.0065
& 7.040
& -- \\
\bottomrule
\end{tabular}
\endgroup
\end{table}

All four block configurations have $\widehat R<1.01$ on
nonconstant link flows. Acceptance depends on the block partition,
with higher rates observed for the finer partitions used on Sioux Falls
and Chicago Sketch. On Chicago Sketch, OD blocking increases the
minimum ESS/1,000 by 6.4\% relative to balanced-origin blocking, but
requires nearly eight times as much time per sweep. The partition changes mixing and cost but not the target law (\cref{prop:block-mh}).

Using the minimum link-flow bulk ESS per unit of total run time, block MH is 12.8 and 76.1 times as efficient as traveler-level Gibbs sampling on Nguyen--Dupuis and Sioux Falls, respectively. The gain comes from the cost of a sweep, which is 57 and 126 times lower for the block sampler (\cref{tab:exp-performance}); per retained draw, traveler-level Gibbs mixes at least as well, and its link-flow draws are nearly uncorrelated from sweep to sweep on both networks. A comparison of the retained Sioux Falls samples gives maximum relative differences of 0.026\% in link-flow means and 6.4\% in standard deviations, using the pooled Gibbs estimates as denominators. With a bulk ESS of at least 1,270 on every link in both samples, the relative Monte Carlo standard error of each estimated standard deviation is about 2\% per sample, so a largest difference of 6.4\% over 76 links is consistent with Monte Carlo error.

\FloatBarrier

\subsection{What the CSUE law adds to SUE: Sioux Falls}
\label{subsec:exp-law}

We examine how route-choice precision $\theta$ and flow quantum
$\varepsilon$ affect the distribution of physical link flows on Sioux
Falls, holding link-cost functions and bushes fixed. Physical OD demands
are preserved for $\varepsilon\le100$ and rounded for larger quanta (\cref{app:implementation}). Each configuration uses four OD-block MH chains with
at most 1,000 decision units per block. For $\varepsilon\le100$, we retain
2,000 sweeps per chain after discarding 100; for $\varepsilon=200,500,1000$,
we retain 4,000 after discarding 1,000.
Across these configurations, the maximum link-flow $\widehat R$ is
below 1.008 and the minimum bulk ESS exceeds 500.

\begin{figure}[tbp]
\centering
\captionsetup{font={small,stretch=1.0}}
\includegraphics[width=1\linewidth]{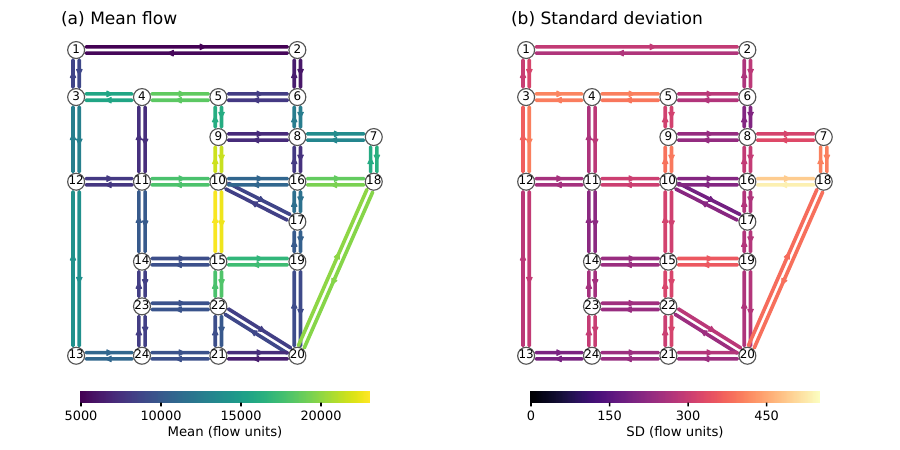}
\caption{Sioux Falls at $\theta=0.3$ and $\varepsilon=100$:
(a) mean physical link flow and (b) its standard deviation.
Parallel arrows distinguish opposite directions; circles identify
nodes. Coordinates are schematic and the color scales are separate.}
\label{fig:sioux-link-marginals}
\end{figure}

\paragraph{Spatial scale of the fluctuations.}
\Cref{fig:sioux-link-marginals} shows the mean and standard deviation
at $\theta=0.3$ and $\varepsilon=100$. This is the largest common
quantum that preserves every Sioux Falls OD demand exactly: the
360,600 physical flow units correspond to 3,606 decision units,
without OD rounding. Mean link flows range from 5,048 to 22,845, and
the largest standard deviation is 533.9 physical flow units, compared
with 54.0 at $\varepsilon=1$. The largest standard deviation occurs on
$18\to16$, whereas the largest mean occurs on $15\to10$.
The links carrying the most traffic are not necessarily those
with the largest fluctuations, so marginal standard deviations add
information that a deterministic flow vector does not contain.

\begin{figure}[tbp]
\centering
\captionsetup{font={small,stretch=1.0}}
\includegraphics[width=\linewidth]{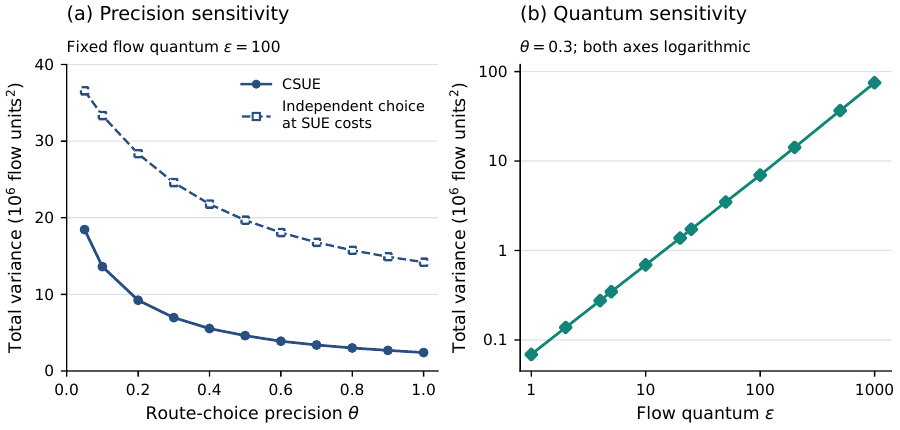}
\caption{Sensitivity of total physical link-flow variance on Sioux Falls.
(a) Precision scan at fixed $\varepsilon=100$: blue circles show CSUE
variance and open squares show the independent-choice reference at the
corresponding SUE costs. (b) Quantum scan at fixed $\theta=0.3$: green
diamonds show CSUE variance for $\varepsilon=1$--$1000$.
Panel (a) uses linear axes and panel (b) uses logarithmic axes.
Both vertical axes report total variance in millions of squared
physical flow units.}
\label{fig:sioux-theta-sensitivity}
\end{figure}

\paragraph{Precision and congestion feedback.}
\Cref{fig:sioux-theta-sensitivity}(a) shows the precision scan
at fixed $\varepsilon=100$. As $\theta$
increases from 0.05 to 1, the sum of marginal physical-flow variances,
$V_{\theta,\varepsilon}=\sum_{a\in\A}\operatorname{Var}(X_a)$,
decreases from $1.846\times10^7$ to $2.404\times10^6$, a factor of 7.7.
Here $X_a$ denotes physical flow, equal to $\varepsilon$ times the
number of decision units using link $a$.
To distinguish frozen-cost choice dispersion from congestion feedback,
we also calculate the independent-choice reference $V^{\mathrm{ind}}_{\theta,\varepsilon}
  =\varepsilon^2\sum_a\sum_w n_w p_{wa}(1-p_{wa})$,
where $n_w=q_w/\varepsilon$ is the number of decision units for OD
pair $w$, and $p_{wa}$ is its probability of using link $a$ under
independent logit choice at the corresponding SUE costs on the same
bushes. Both variances decrease as precision increases, but the CSUE
variance is lower throughout the scan. At $\theta=0.3$, it is only 28.3\%
of the independent-choice value. Independent choice at SUE costs thus
overstates total variability, because it omits the congestion feedback
that damps fluctuations.

\paragraph{Flow quantum and dispersion.}
\Cref{fig:sioux-theta-sensitivity}(b) shows the quantum scan at
fixed $\theta=0.3$, using
$\varepsilon\in\{1,\allowbreak 2,\allowbreak 4,\allowbreak 5,\allowbreak
10,\allowbreak 20,\allowbreak 25,\allowbreak 50,\allowbreak 100,\allowbreak
200,\allowbreak 500,\allowbreak 1000\}$.
Larger quanta use rounded demands (\cref{app:implementation}).

Total physical variance increases from $6.893\times10^4$ at
$\varepsilon=1$ to $7.504\times10^7$ at $\varepsilon=1000$.
The curve is approximately straight on the log--log axes, with a
slope close to one: physical-flow variance grows roughly in proportion
to the flow quantum over the tested range, as anticipated in \cref{subsec:beckmann}.

\FloatBarrier

\subsection{Applications on Chicago Sketch: sensor assessment and origin attribution}
\label{subsec:exp-sensors}

The layered CSUE law supports two complementary uses of traffic-flow samples. Covariances quantify how a count on one link informs flows elsewhere, while the retained origin labels identify the sources of traffic on a link and the variability of that composition. We illustrate both uses on Chicago Sketch. Both examples condition on known OD demand, cost functions, and bushes, and the sensor example assumes noiseless counts. Both examples use $\theta=0.3$.

\subsubsection{Covariance-based assessment of sensor locations}

We consider counts on directed physical links, with centroid connectors excluded from the 2,176 candidate locations. The sensor calculation uses covariance estimates from sweeps 201--1600 of four chains for each of the origin-block and OD-block samplers. A sensor on link $a$ observes total flow $X_a=\sum_o X_{a,o}$, where $X_{a,o}=\sum_{g:\,o_g=o}X_a^g$ is the contribution from origin $o$. Covariances allow these total counts to inform origin contributions on both observed and unobserved links.

Let $\mathbf Y=(X_{a_j,o_j})_{j=1}^d$ collect the origin--link flows to be predicted and $\mathbf Z_S=(X_a)_{a\in S}$ the total flows counted on a sensor set $S$. Using $\widehat\Sigma_{UV}$ for the estimated cross-covariance of vectors $U$ and $V$, the weighted residual variance of the best affine predictor is 
\[
J_{\mathrm O}(S)
 =\operatorname{tr}\!\left[
 W\left(\widehat\Sigma_{YY}
   -\widehat\Sigma_{Y Z_S}\widehat\Sigma_{Z_S Z_S}^{\dagger}
    \widehat\Sigma_{Z_S Y}\right)\right].
\]
Here $W=\diag(w_1,\ldots,w_d)$ assigns nonnegative weights to the prediction errors. We use square-root flow weights $w_j=\sqrt{\widehat\mu_j}/\sum_{k=1}^d\sqrt{\widehat\mu_k}$, where $\widehat\mu_j$ is the sample mean of $Y_j$; busier origin contributions thus receive greater weight. The Moore--Penrose inverse $\dagger$ allows linearly dependent counts. 

\begin{figure}[!htbp]
\centering
\captionsetup{font={small,stretch=1.0}}
\includegraphics[width=\linewidth]{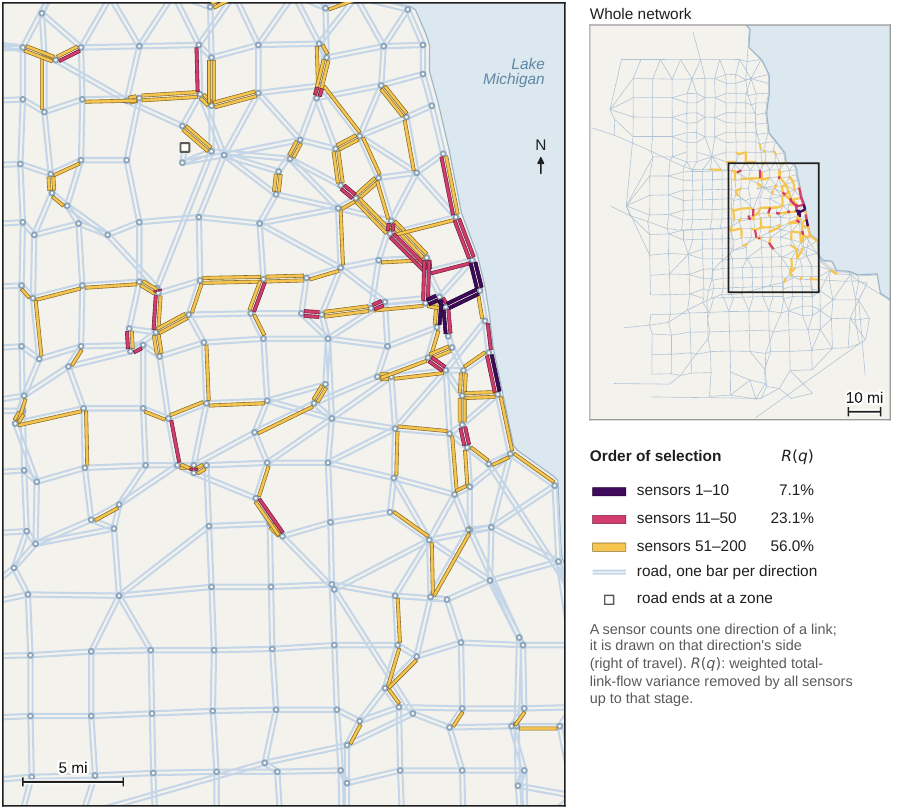}
\caption{Nested sensor sets on Chicago Sketch at $\theta=0.3$. Central enlargement (left) and whole-network context (upper right); the box in the whole-network view marks the extent of the enlargement. Purple, rose, and yellow indicate sensors added at ranks 1–10, 11–50, and 51–200, respectively. Percentages are cumulative origin-level weighted variance reductions $R_{\mathrm O}(q)$ relative to the same no-count baseline. In the enlargement, each directed link is drawn on the right-hand side of its direction of travel; a colored bar marks the counted direction.}
\label{fig:chicago-sensor-batches}
\end{figure}

Starting from an empty set, we sequentially add the candidate link that gives the largest reduction in $J_{\mathrm O}$ given the links already selected. The sets $S_q$ of $q$ counted links in \cref{fig:chicago-sensor-batches} are nested: $S_{10}\subset S_{50}\subset S_{200}$. For $J_{\mathrm O}(\varnothing)>0$, their cumulative variance reduction relative to the same no-count baseline is
\begin{equation}
  R_{\mathrm O}(q)
  =100\left[1-\frac{J_{\mathrm O}(S_q)}{J_{\mathrm O}(\varnothing)}\right].
\end{equation}

\Cref{fig:chicago-sensor-batches} shows cumulative reductions of 7.1\%, 23.1\%, and 56.0\% at budgets of 10, 50, and 200 sensors, respectively. The colors distinguish the links added at each stage; each percentage includes all sensors retained up to that budget. With 200 sensors, 9.2\% of the candidate links are observed and 44.0\% of the initial weighted variance remains.

The map shows early selections concentrated along a few central and eastern corridors, followed by broader coverage as the budget increases.

\FloatBarrier
\subsubsection{Origin contributions and their variability}

\begin{figure}[!htbp]
\centering
\captionsetup{font={small,stretch=1.0}}
\includegraphics[width=\linewidth]{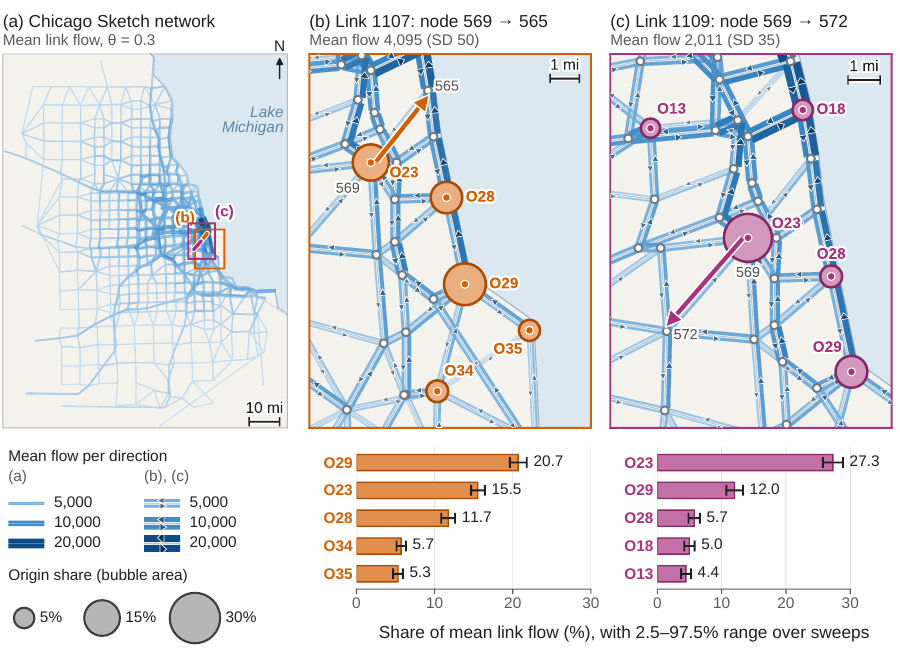}
\caption{Origin attribution on Chicago Sketch at $\theta=0.3$. (a) Whole-network context; (b) link 1107 ($569\rightarrow565$); and (c) link 1109 ($569\rightarrow572$). Road width and shade show mean flow in each direction; the whole-network panel and the close-ups use separate width scales, as shown in the legend. Coloured arrows mark the two selected links. Bubbles are centered at the network access nodes of the five leading origins; O23, for example, denotes origin zone 23. Bubble area and bar length represent $100s_{a,o}$, without renormalizing over the displayed origins. All bubbles use the same area scale. Error bars show the 2.5--97.5\% quantiles of per-sweep shares across 2,000 retained draws.}
\label{fig:chicago-link-attribution}
\end{figure}

The retained layers also describe traffic composition. For link $a$ and origin $o$, let $s_{a,o}=\E[X_{a,o}]/\E[X_a]$ be the share of mean flow contributed by origin $o$; because $X_{a,o}$ sums all blocks of origin $o$, the definition applies to any blocking. \Cref{fig:chicago-link-attribution} uses the four balanced-origin chains at $\theta=0.3$ in \cref{tab:exp-performance} (sweeps 101--600 of each chain, 2,000 pooled draws) and shows two central links with mean flow above 2,000 that leave node 569. Their leading origins differ: O29 contributes 20.7\% of mean flow on link 1107, and O23 contributes 27.3\% on link 1109. The error bars, the 2.5--97.5\% ranges of the per-sweep shares $X_{a,o}^{(t)}/X_a^{(t)}$, show how much the composition fluctuates around the single composition that a deterministic assignment supplies.

\FloatBarrier

\section{Conclusion}
\label{sec:discussion}

This paper develops an explicit Gibbs representation and a path-free
sampling method for finite-population logit CSUE. Under separable
costs and a common precision, the leave-one-out conditionals determine
a unique Gibbs law with a left-endpoint discrete congestion potential, and on
fixed acyclic bushes origin- or OD-layered flows admit an explicit
unnormalized law with product-form multiplicities. Integer multinomial
loading and a layer-based Metropolis--Hastings correction preserve
this law without enumerating paths or storing individual routes.

The value of CSUE lies in the joint variability of flows rather than in a better point estimate: deterministic logit SUE is recovered in the nonatomic limit and is close to the finite-population mean in the enumerable example. The validation examples show that this variability cannot be obtained by loading travelers independently at frozen costs, which leaves mean flows nearly unchanged but overstates standard deviations and distorts correlations. On Sioux Falls, the sensitivity scans separate the effects of route-choice precision and decision quantum on total variance, and on Chicago Sketch the sampled covariance ranks traffic-count locations by the variance they remove.

Computationally, retaining layers avoids path enumeration without approximating the CSUE law, and on fixed bushes every block partition targets the same total-flow law. The cost of a sweep is governed by bush sizes and the number of blocks. With a fixed number of mean-field iterations, large blocks may lose acceptance as the population grows (\cref{rem:acceptance-scaling}), and the rejection rate, which brackets the conditional proposal error (\cref{prop:block-error-bound}), indicates which blocks to split.

The analysis has limitations. The bush route sets are part of the model, so sensitivity to their construction is a modeling question that the MH correction does not address. The Gibbs representation requires separable costs and a common precision (\cref{rem:logit-dynamics}); heterogeneous congestion sensitivities, nonseparable interactions, and day-to-day dynamics are outside the model. The experiments condition on known demand and, for the sensor example, on noiseless counts, so measurement noise, demand variation, and validation against observed counts remain to be added.

The forward law is the latent assignment model for Bayesian inference from traffic counts. Combined with an observation model for counts on a subset of links and with priors on the precision, the link-cost parameters, and the OD demands, it gives a joint posterior over layers and parameters. For known parameters and demand, the CSUE normalizing constant
cancels from latent-flow MH ratios, and the block update needs only an additional likelihood ratio in its acceptance probability. Inference for unknown precision
or cost parameters must account for its parameter dependence
\citep{Younes1988,GeyerThompson1992}.
Exchange methods \citep{MurrayGhahramaniMacKay2006} are a candidate, but they require independent (perfect) draws from the model, which an MCMC sampler does not provide. Unknown demand
also changes the feasible flow space and multiplicities.
Finally, a nonuniform reference measure on routes would give weighted logit models such as path-size logit \citep{BenAkivaBierlaire1999}; path-free sampling then requires route weights that factor over links.

\section*{Acknowledgements}

This research was supported by the Natural Sciences and Engineering Research Council of Canada (NSERC) through a Discovery Grant (RGPIN-2025-04479).

\section*{Declaration of generative AI and AI-assisted technologies in the writing process}

During the preparation of this work, the authors used generative AI tools (OpenAI's ChatGPT and Anthropic's Claude) to edit and proofread the text, to locate relevant literature, and to check the arithmetic of the illustrative examples. The authors verified every reference against its original source and checked every numerical result independently. The conception of the paper, its structure and arguments, and all final editorial decisions are the authors' own, and the authors take full responsibility for the content of the publication.

\bibliography{references}

\begin{thebibliography}{58}
\providecommand{\natexlab}[1]{#1}
\providecommand{\url}[1]{\texttt{#1}}
\expandafter\ifx\csname urlstyle\endcsname\relax
  \providecommand{\doi}[1]{doi: #1}\else
  \providecommand{\doi}{doi: \begingroup \urlstyle{rm}\Url}\fi

\bibitem[Ahuja et~al.(1993)Ahuja, Magnanti, and Orlin]{AhujaMagnantiOrlin1993}
Ravindra~K. Ahuja, Thomas~L. Magnanti, and James~B. Orlin.
\newblock \emph{Network Flows: Theory, Algorithms, and Applications}.
\newblock Prentice Hall, Englewood Cliffs, NJ, 1993.

\bibitem[Akamatsu(1996)]{Akamatsu1996}
Takashi Akamatsu.
\newblock Cyclic flows, {M}arkov process and stochastic traffic assignment.
\newblock \emph{Transportation Research Part B: Methodological}, 30\penalty0
  (5):\penalty0 369--386, 1996.

\bibitem[Akamatsu(1997)]{Akamatsu1997}
Takashi Akamatsu.
\newblock Decomposition of path choice entropy in general transport networks.
\newblock \emph{Transportation Science}, 31\penalty0 (4):\penalty0 349--362,
  1997.

\bibitem[Arnold and Press(1989)]{ArnoldPress1989}
Barry~C. Arnold and S.~James Press.
\newblock Compatible conditional distributions.
\newblock \emph{Journal of the American Statistical Association}, 84\penalty0
  (405):\penalty0 152--156, 1989.

\bibitem[Baillon and Cominetti(2008)]{BaillonCominetti2008}
Jean-Bernard Baillon and Roberto Cominetti.
\newblock {M}arkovian traffic equilibrium.
\newblock \emph{Mathematical Programming}, 111\penalty0 (1--2):\penalty0
  33--56, 2008.

\bibitem[Bar-Gera(2002)]{BarGera2002}
Hillel Bar-Gera.
\newblock Origin-based algorithm for the traffic assignment problem.
\newblock \emph{Transportation Science}, 36\penalty0 (4):\penalty0 398--417,
  2002.

\bibitem[Beckmann et~al.(1956)Beckmann, McGuire, and Winsten]{Beckmann1956}
Martin Beckmann, C.~B. McGuire, and Christopher~B. Winsten.
\newblock \emph{Studies in the Economics of Transportation}.
\newblock Yale University Press, New Haven, CT, 1956.

\bibitem[Bell(1995)]{Bell1995}
Michael G.~H. Bell.
\newblock Alternatives to {D}ial's logit assignment algorithm.
\newblock \emph{Transportation Research Part B: Methodological}, 29\penalty0
  (4):\penalty0 287--295, 1995.

\bibitem[Ben-Akiva and Bierlaire(1999)]{BenAkivaBierlaire1999}
Moshe Ben-Akiva and Michel Bierlaire.
\newblock Discrete choice methods and their applications to short term travel
  decisions.
\newblock In Randolph~W. Hall, editor, \emph{Handbook of Transportation
  Science}, volume~23 of \emph{International Series in Operations Research \&
  Management Science}, pages 5--33. Kluwer, Boston, MA, 1999.

\bibitem[Besag(1974)]{Besag1974}
Julian Besag.
\newblock Spatial interaction and the statistical analysis of lattice systems.
\newblock \emph{Journal of the Royal Statistical Society: Series B
  (Methodological)}, 36\penalty0 (2):\penalty0 192--236, 1974.

\bibitem[Blume(1993)]{Blume1993}
Lawrence~E. Blume.
\newblock The statistical mechanics of strategic interaction.
\newblock \emph{Games and Economic Behavior}, 5\penalty0 (3):\penalty0
  387--424, 1993.

\bibitem[Brook(1964)]{Brook1964}
D.~Brook.
\newblock On the distinction between the conditional probability and the joint
  probability approaches in the specification of nearest-neighbour systems.
\newblock \emph{Biometrika}, 51\penalty0 (3/4):\penalty0 481--483, 1964.

\bibitem[Cantarella and Cascetta(1995)]{CantarellaCascetta1995}
Giulio~E. Cantarella and Ennio Cascetta.
\newblock Dynamic processes and equilibrium in transportation networks: Towards
  a unifying theory.
\newblock \emph{Transportation Science}, 29\penalty0 (4):\penalty0 305--329,
  1995.

\bibitem[Cascetta(1989)]{Cascetta1989}
Ennio Cascetta.
\newblock A stochastic process approach to the analysis of temporal dynamics in
  transportation networks.
\newblock \emph{Transportation Research Part B: Methodological}, 23\penalty0
  (1):\penalty0 1--17, 1989.

\bibitem[Clark and Watling(2005)]{ClarkWatling2005}
Stephen Clark and David Watling.
\newblock Modelling network travel time reliability under stochastic demand.
\newblock \emph{Transportation Research Part B: Methodological}, 39\penalty0
  (2):\penalty0 119--140, 2005.

\bibitem[Daganzo and Sheffi(1977)]{DaganzoSheffi1977}
Carlos~F. Daganzo and Yosef Sheffi.
\newblock On stochastic models of traffic assignment.
\newblock \emph{Transportation Science}, 11\penalty0 (3):\penalty0 253--274,
  1977.

\bibitem[Davis and Nihan(1993)]{DavisNihan1993}
Gary~A. Davis and Nancy~L. Nihan.
\newblock Large population approximations of a general stochastic traffic
  assignment model.
\newblock \emph{Operations Research}, 41\penalty0 (1):\penalty0 169--178, 1993.

\bibitem[Dial(1971)]{Dial1971}
Robert~B. Dial.
\newblock A probabilistic multipath traffic assignment model which obviates
  path enumeration.
\newblock \emph{Transportation Research}, 5\penalty0 (2):\penalty0 83--111,
  1971.

\bibitem[Dial(2006)]{Dial2006}
Robert~B. Dial.
\newblock A path-based user-equilibrium traffic assignment algorithm that
  obviates path storage and enumeration.
\newblock \emph{Transportation Research Part B: Methodological}, 40\penalty0
  (10):\penalty0 917--936, 2006.

\bibitem[Fisk(1980)]{Fisk1980}
Caroline Fisk.
\newblock Some developments in equilibrium traffic assignment.
\newblock \emph{Transportation Research Part B: Methodological}, 14\penalty0
  (3):\penalty0 243--255, 1980.

\bibitem[Fosgerau et~al.(2013)Fosgerau, Frejinger, and
  Karlstr{\"o}m]{FosgerauFrejingerKarlstrom2013}
Mogens Fosgerau, Emma Frejinger, and Anders Karlstr{\"o}m.
\newblock A link based network route choice model with unrestricted choice set.
\newblock \emph{Transportation Research Part B: Methodological}, 56:\penalty0
  70--80, 2013.

\bibitem[Gentili and Mirchandani(2012)]{GentiliMirchandani2012}
Monica Gentili and Pitu~B. Mirchandani.
\newblock Locating sensors on traffic networks: Models, challenges and research
  opportunities.
\newblock \emph{Transportation Research Part C: Emerging Technologies},
  24:\penalty0 227--255, 2012.

\bibitem[Geyer and Thompson(1992)]{GeyerThompson1992}
Charles~J. Geyer and Elizabeth~A. Thompson.
\newblock Constrained {M}onte {C}arlo maximum likelihood for dependent data.
\newblock \emph{Journal of the Royal Statistical Society: Series B
  (Methodological)}, 54\penalty0 (3):\penalty0 657--699, 1992.

\bibitem[Hastings(1970)]{Hastings1970}
W.~K. Hastings.
\newblock {M}onte {C}arlo sampling methods using {M}arkov chains and their
  applications.
\newblock \emph{Biometrika}, 57\penalty0 (1):\penalty0 97--109, 1970.

\bibitem[Hazelton(1998)]{Hazelton1998}
Martin~L. Hazelton.
\newblock Some remarks on stochastic user equilibrium.
\newblock \emph{Transportation Research Part B: Methodological}, 32\penalty0
  (2):\penalty0 101--108, 1998.

\bibitem[Hazelton(2000)]{Hazelton2000}
Martin~L. Hazelton.
\newblock Estimation of origin--destination matrices from link flows on
  uncongested networks.
\newblock \emph{Transportation Research Part B: Methodological}, 34\penalty0
  (7):\penalty0 549--566, 2000.

\bibitem[Hazelton(2008)]{Hazelton2008}
Martin~L. Hazelton.
\newblock Statistical inference for time varying origin--destination matrices.
\newblock \emph{Transportation Research Part B: Methodological}, 42\penalty0
  (6):\penalty0 542--552, 2008.

\bibitem[Hazelton(2015)]{Hazelton2015}
Martin~L. Hazelton.
\newblock Network tomography for integer-valued traffic.
\newblock \emph{The Annals of Applied Statistics}, 9\penalty0 (1):\penalty0
  474--506, 2015.

\bibitem[Hazelton(2022)]{Hazelton2022}
Martin~L. Hazelton.
\newblock The emergence of stochastic user equilibria in day-to-day traffic
  models.
\newblock \emph{Transportation Research Part B: Methodological}, 158:\penalty0
  102--112, 2022.

\bibitem[Hazelton and Najim(2024)]{HazeltonNajim2024}
Martin~L. Hazelton and Lara Najim.
\newblock Using traffic assignment models to assist {B}ayesian inference for
  origin--destination matrices.
\newblock \emph{Transportation Research Part B: Methodological}, 186:\penalty0
  103019, 2024.

\bibitem[Hazelton and Watling(2004)]{HazeltonWatling2004}
Martin~L. Hazelton and David~P. Watling.
\newblock Computation of equilibrium distributions of {M}arkov
  traffic-assignment models.
\newblock \emph{Transportation Science}, 38\penalty0 (3):\penalty0 331--342,
  2004.

\bibitem[Hazelton et~al.(1996)Hazelton, Lee, and Polak]{HazeltonLeePolak1996}
Martin~L. Hazelton, Seungjae Lee, and John~W. Polak.
\newblock Stationary states in stochastic process models of traffic assignment:
  A {M}arkov chain {M}onte {C}arlo approach.
\newblock In Jean-Baptiste Lesort, editor, \emph{Transportation and Traffic
  Theory: Proceedings of the 13th International Symposium on Transportation and
  Traffic Theory}, pages 341--357. Pergamon, Oxford, 1996.

\bibitem[Kleer(2023)]{Kleer2023}
Pieter Kleer.
\newblock Sampling from the {Gibbs} distribution in congestion games.
\newblock \emph{Mathematics of Operations Research}, 48\penalty0 (4):\penalty0
  1846--1870, 2023.

\bibitem[Krause et~al.(2008)Krause, Singh, and
  Guestrin]{KrauseSinghGuestrin2008}
Andreas Krause, Ajit Singh, and Carlos Guestrin.
\newblock Near-optimal sensor placements in {G}aussian processes: Theory,
  efficient algorithms and empirical studies.
\newblock \emph{Journal of Machine Learning Research}, 9:\penalty0 235--284,
  2008.

\bibitem[LeBlanc et~al.(1975)LeBlanc, Morlok, and
  Pierskalla]{LeBlancMorlokPierskalla1975}
Larry~J. LeBlanc, Edward~K. Morlok, and William~P. Pierskalla.
\newblock An efficient approach to solving the road network equilibrium traffic
  assignment problem.
\newblock \emph{Transportation Research}, 9\penalty0 (5):\penalty0 309--318,
  1975.

\bibitem[Maher(1983)]{Maher1983}
M.~J. Maher.
\newblock Inferences on trip matrices from observations on link volumes: A
  {B}ayesian statistical approach.
\newblock \emph{Transportation Research Part B: Methodological}, 17\penalty0
  (6):\penalty0 435--447, 1983.

\bibitem[Monderer and Shapley(1996)]{MondererShapley1996}
Dov Monderer and Lloyd~S. Shapley.
\newblock Potential games.
\newblock \emph{Games and Economic Behavior}, 14\penalty0 (1):\penalty0
  124--143, 1996.

\bibitem[Murray et~al.(2006)Murray, Ghahramani, and
  MacKay]{MurrayGhahramaniMacKay2006}
Iain Murray, Zoubin Ghahramani, and David J.~C. MacKay.
\newblock {MCMC} for doubly-intractable distributions.
\newblock In \emph{Proceedings of the 22nd Annual Conference on Uncertainty in
  Artificial Intelligence}, pages 359--366. AUAI Press, 2006.

\bibitem[Nakayama and Watling(2014)]{NakayamaWatling2014}
Shoichiro Nakayama and David Watling.
\newblock Consistent formulation of network equilibrium with stochastic flows.
\newblock \emph{Transportation Research Part B: Methodological}, 66:\penalty0
  50--69, 2014.

\bibitem[Nguyen and Dupuis(1984)]{NguyenDupuis1984}
Sang Nguyen and Clermont Dupuis.
\newblock An efficient method for computing traffic equilibria in networks with
  asymmetric transportation costs.
\newblock \emph{Transportation Science}, 18\penalty0 (2):\penalty0 185--202,
  1984.

\bibitem[Nie(2010)]{Nie2010}
Yu~(Marco) Nie.
\newblock A class of bush-based algorithms for the traffic assignment problem.
\newblock \emph{Transportation Research Part B: Methodological}, 44\penalty0
  (1):\penalty0 73--89, 2010.

\bibitem[Parry and Hazelton(2013)]{ParryHazelton2013}
Katharina Parry and Martin~L. Hazelton.
\newblock {B}ayesian inference for day-to-day dynamic traffic models.
\newblock \emph{Transportation Research Part B: Methodological}, 50:\penalty0
  104--115, 2013.

\bibitem[Rosenthal(1973{\natexlab{a}})]{Rosenthal1973}
Robert~W. Rosenthal.
\newblock The network equilibrium problem in integers.
\newblock \emph{Networks}, 3\penalty0 (1):\penalty0 53--59, 1973{\natexlab{a}}.

\bibitem[Rosenthal(1973{\natexlab{b}})]{Rosenthal1973IJGT}
Robert~W. Rosenthal.
\newblock A class of games possessing pure-strategy {N}ash equilibria.
\newblock \emph{International Journal of Game Theory}, 2\penalty0 (1):\penalty0
  65--67, 1973{\natexlab{b}}.

\bibitem[Sandholm(2010)]{Sandholm2010}
William~H. Sandholm.
\newblock \emph{Population Games and Evolutionary Dynamics}.
\newblock MIT Press, Cambridge, MA, 2010.

\bibitem[Sandholm and Staudigl(2018)]{SandholmStaudigl2018}
William~H. Sandholm and Mathias Staudigl.
\newblock Sample path large deviations for stochastic evolutionary game
  dynamics.
\newblock \emph{Mathematics of Operations Research}, 43\penalty0 (4):\penalty0
  1348--1377, 2018.

\bibitem[Sheffi(1985)]{Sheffi1985}
Yosef Sheffi.
\newblock \emph{Urban Transportation Networks: Equilibrium Analysis with
  Mathematical Programming Methods}.
\newblock Prentice-Hall, Englewood Cliffs, NJ, 1985.

\bibitem[Tebaldi and West(1998)]{TebaldiWest1998}
Claudia Tebaldi and Mike West.
\newblock {B}ayesian inference on network traffic using link count data.
\newblock \emph{Journal of the American Statistical Association}, 93\penalty0
  (442):\penalty0 557--573, 1998.

\bibitem[Tierney(1994)]{Tierney1994}
Luke Tierney.
\newblock {M}arkov chains for exploring posterior distributions.
\newblock \emph{The Annals of Statistics}, 22\penalty0 (4):\penalty0
  1701--1728, 1994.

\bibitem[{Transportation Networks for Research Core Team}(2026)]{TNTP}
{Transportation Networks for Research Core Team}.
\newblock Transportation networks for research.
\newblock \url{https://github.com/bstabler/TransportationNetworks}, 2026.
\newblock Accessed November 2026.

\bibitem[Vehtari et~al.(2021)Vehtari, Gelman, Simpson, Carpenter, and
  B{\"u}rkner]{Vehtari2021}
Aki Vehtari, Andrew Gelman, Daniel Simpson, Bob Carpenter, and Paul-Christian
  B{\"u}rkner.
\newblock Rank-normalization, folding, and localization: An improved
  {$\widehat{R}$} for assessing convergence of {MCMC}.
\newblock \emph{Bayesian Analysis}, 16\penalty0 (2):\penalty0 667--718, 2021.

\bibitem[Wardrop(1952)]{Wardrop1952}
J.~G. Wardrop.
\newblock {S}ome theoretical aspects of road traffic research.
\newblock \emph{Proceedings of the Institution of Civil Engineers}, 1\penalty0
  (3):\penalty0 325--362, 1952.

\bibitem[Watling(1996)]{Watling1996}
David Watling.
\newblock Asymmetric problems and stochastic process models of traffic
  assignment.
\newblock \emph{Transportation Research Part B: Methodological}, 30\penalty0
  (5):\penalty0 339--357, 1996.

\bibitem[Watling(2002)]{Watling2002}
David~P. Watling.
\newblock A second order stochastic network equilibrium model, {I}: Theoretical
  foundation.
\newblock \emph{Transportation Science}, 36\penalty0 (2):\penalty0 149--166,
  2002.

\bibitem[Watling and Cantarella(2013)]{WatlingCantarella2013}
David~P. Watling and Giulio~E. Cantarella.
\newblock Modelling sources of variation in transportation systems: Theoretical
  foundations of day-to-day dynamic models.
\newblock \emph{Transportmetrica B: Transport Dynamics}, 1\penalty0
  (1):\penalty0 3--32, 2013.

\bibitem[Watling and Hazelton(2018)]{WatlingHazelton2018}
David~P. Watling and Martin~L. Hazelton.
\newblock Asymptotic approximations of transient behaviour for day-to-day
  traffic models.
\newblock \emph{Transportation Research Part B: Methodological}, 118:\penalty0
  90--105, 2018.

\bibitem[Yang and Zhou(1998)]{YangZhou1998}
Hai Yang and Jing Zhou.
\newblock Optimal traffic counting locations for origin--destination matrix
  estimation.
\newblock \emph{Transportation Research Part B: Methodological}, 32\penalty0
  (2):\penalty0 109--126, 1998.

\bibitem[Younes(1988)]{Younes1988}
Laurent Younes.
\newblock Estimation and annealing for {G}ibbsian fields.
\newblock \emph{Annales de l'Institut Henri Poincar{\'e}, Probabilit{\'e}s et
  Statistiques}, 24\penalty0 (2):\penalty0 269--294, 1988.

\end{thebibliography}

\clearpage
\appendix
\crefalias{section}{appendix}
\crefalias{subsection}{subappendix}

\section{Proofs and Supplementary Results}
\label{app:proofs}

\subsection{Beckmann limit of the discrete potential}
\label{app:beckmann-limit}

\begin{lemma}[Beckmann limit]
\label{prop:beckmann-limit}
Suppose every link-cost function $c_a$ is continuous.
For any sequence of lattice link flows
$\bx^\varepsilon\to\bx$ as $\varepsilon\downarrow0$,
with $x_a^\varepsilon/\varepsilon\in\Nzero$
for every link,
\begin{equation}
    \varepsilon\Phi_\varepsilon(\bx^\varepsilon)
    \longrightarrow B(\bx).
    \label{eq:beckmann-limit}
\end{equation}
If every $c_a$ is also nondecreasing, then for any
$\varepsilon>0$ and any lattice flow $\bx$ with
$x_a/\varepsilon\in\Nzero$ on every link,
\begin{equation}
    0
    \le
    B(\bx)-\varepsilon\Phi_\varepsilon(\bx)
    \le
    \varepsilon
    \sum_{a\in\A}
    \bigl[c_a(x_a)-c_a(0)\bigr].
    \label{eq:beckmann-bound}
\end{equation}
\end{lemma}

\begin{proof}
Since $\bx^\varepsilon$ converges, $\bar x:=\sup_{\varepsilon}\max_{a\in\A}x_a^\varepsilon<\infty$. Let $\varpi_a$ be the modulus of continuity of $c_a$ on $[0,\bar x]$. Then
\[
  \bigl|\varepsilon\Phi_\varepsilon(\bx^\varepsilon)-B(\bx^\varepsilon)\bigr|
  =\Bigl|\sum_{a\in\A}\sum_{m=0}^{x_a^\varepsilon/\varepsilon-1}
    \int_{m\varepsilon}^{(m+1)\varepsilon}
    \bigl[c_a(m\varepsilon)-c_a(u)\bigr]\,du\Bigr|
  \le\sum_{a\in\A}x_a^\varepsilon\,\varpi_a(\varepsilon)
  \le\bar x\sum_{a\in\A}\varpi_a(\varepsilon),
\]
which tends to zero because each $c_a$ is uniformly continuous on $[0,\bar x]$. Continuity of $B$ gives $B(\bx^\varepsilon)\to B(\bx)$, which proves Equation~\eqref{eq:beckmann-limit}.

If $c_a$ is nondecreasing, then $c_a(m\varepsilon)\le c_a(u)\le c_a((m+1)\varepsilon)$ for $u\in[m\varepsilon,(m+1)\varepsilon]$. Each link integral therefore lies between its left- and right-endpoint Riemann sums, whose difference telescopes to $\varepsilon[c_a(x_a)-c_a(0)]$. Summing over links gives Equation~\eqref{eq:beckmann-bound}.
\end{proof}%

\subsection{Heterogeneous precision}
\label{app:heterogeneous}

Consider two travelers of the same OD pair choosing between two parallel links $1$ and $2$, with precisions $\theta_1$ and $\theta_2$ in Equation~\eqref{eq:csue-conditional}, and write $(r_1,r_2)$ for a route profile. Any joint distribution with these conditionals has ratios along the cycle $(1,1)\to(2,1)\to(2,2)\to(1,2)\to(1,1)$ whose product is one. Each ratio is a conditional odds of one traveler with the other held fixed, and the product of the four conditional odds equals
\[
  \exp\bigl\{-(\theta_1-\theta_2)\,[c_1(1)-c_1(0)+c_2(1)-c_2(0)]\bigr\}.
\]
When $\theta_1\ne\theta_2$ and the bracket is nonzero, as for strictly increasing costs, no joint distribution has these conditionals.

\subsection{Proof of \texorpdfstring{\cref{thm:large-population-sue}}{the large-population limit}}
\label{app:large-population-details}

Write $n_w^\varepsilon=q_w/\varepsilon$ and $N_{wr}^\varepsilon=f_{wr}^\varepsilon/\varepsilon$, and let $\mathcal F_\varepsilon(\mathbf q):=\{\bfv\in\mathcal F(\mathbf q):f_{wr}/\varepsilon\in\Nzero\text{ for all }w,r\}$ be the feasible lattice. By \cref{thm:gibbs}, applied to the scaled system as noted in \cref{subsec:beckmann}, $\pi_F^\varepsilon(\bfv^\varepsilon)=\Psi_\varepsilon(\bfv^\varepsilon)/Z_F^\varepsilon$ with the unnormalized weight
\begin{equation}
    \Psi_\varepsilon(\bfv^\varepsilon)
    =
    \exp\!\left\{
        -\theta
        \Phi_\varepsilon(\Delta\bfv^\varepsilon)
    \right\}
    \prod_{w\in\W}
    \frac{
        n_w^\varepsilon!
    }{
        \prod_{r\in\Rset_w}
        N_{wr}^\varepsilon!
    }.
\end{equation}

\paragraph{Rate of the unnormalized weight.}
By \cref{prop:beckmann-limit}, $\varepsilon\Phi_\varepsilon(\Delta\bfv^\varepsilon)\to B(\Delta\bfv)$ for any $\bfv^\varepsilon\to\bfv$. For the multiplicity, Stirling's formula may be written as $\log(k!)=k\log k-k+O(\log(k+1))$ for integers $k\ge0$, with the convention $0\log0=0$. Since all feasible physical path flows are bounded by the fixed
OD demands, multiplication by $\varepsilon$ makes the remainder
$o(1)$ uniformly over the feasible lattice. Thus
\begin{equation}
    \varepsilon\log\!\left(N_{wr}^\varepsilon!\right)
    =
    f_{wr}^\varepsilon
    \log\!\left(\frac{f_{wr}^\varepsilon}{\varepsilon}\right)
    -f_{wr}^\varepsilon+o(1),
    \qquad
    \varepsilon\log\!\left(n_w^\varepsilon!\right)
    =
    q_w\log\!\left(\frac{q_w}{\varepsilon}\right)
    -q_w+o(1).
\end{equation}
Because $\sum_{r\in\Rset_w}f_{wr}^\varepsilon=q_w$, the $\log\varepsilon$ terms and the linear terms cancel when these expansions are combined, so that
\begin{equation}
    -\varepsilon
    \log
    \left[
        \frac{n_w^\varepsilon!}
        {\prod_{r\in\Rset_w}N_{wr}^\varepsilon!}
    \right]
    =
    \sum_{r\in\Rset_w}
    f_{wr}^\varepsilon
    \log\!\left(\frac{f_{wr}^\varepsilon}{q_w}\right)
    +o(1)
    \longrightarrow
    \sum_{r\in\Rset_w}
    f_{wr}\log\!\left(\frac{f_{wr}}{q_w}\right),
\end{equation}
where the limit uses the continuity of $x\log x$ at zero. Summing over OD pairs and adding the congestion term gives
\begin{equation}
    -\varepsilon\log\Psi_\varepsilon(\bfv^\varepsilon)
    =
    I(\bfv)+o(1),
    \label{eq:app-unnormalized-rate}
\end{equation}
which is Equation~\eqref{eq:large-population-rate} with $\kappa_\varepsilon=\varepsilon\log Z_F^\varepsilon$.

\paragraph{Uniformity.}
The convergence in Equation~\eqref{eq:app-unnormalized-rate} holds along every lattice sequence that converges in $\mathcal F(\mathbf q)$, and $I$ is continuous on the compact set $\mathcal F(\mathbf q)$. Hence
\begin{equation}
  \sup_{\bfv\in\mathcal F_\varepsilon(\mathbf q)}
  \bigl|-\varepsilon\log\Psi_\varepsilon(\bfv)-I(\bfv)\bigr|
  \longrightarrow0.
  \label{eq:app-uniform-rate}
\end{equation}
Otherwise there would be $\delta>0$ and lattice points along a subsequence with error at least $\delta$. By compactness, a further subsequence would converge, contradicting Equation~\eqref{eq:app-unnormalized-rate}.%

\paragraph{Concentration.}
Because all route sets are finite, the feasible lattice has
only polynomially many states in $1/\varepsilon$:
\[
    \left|\mathcal F_\varepsilon(\mathbf q)\right|
    =
    \prod_{w\in\W}
    \binom{n_w^\varepsilon+|\Rset_w|-1}{|\Rset_w|-1},
    \qquad\text{so}\qquad
    \varepsilon\log\left|\mathcal F_\varepsilon(\mathbf q)\right|
    \longrightarrow0.
\]
Let $I_\star=\min_{\bfv\in\mathcal F(\mathbf q)}I(\bfv)$ and let $\bfv^\star$ be its unique minimizer, proved below. For the upper bound, $Z_F^\varepsilon\le|\mathcal F_\varepsilon(\mathbf q)|\max_{\bfv}\Psi_\varepsilon(\bfv)$, so Equation~\eqref{eq:app-uniform-rate} gives $\varepsilon\log Z_F^\varepsilon\le-I_\star+o(1)$. For the lower bound, round all but one coordinate of $\bfv^\star$ within each OD pair down to multiples of $\varepsilon$, and assign the remaining demand to its last route. Since $q_w/\varepsilon$ is an integer, the resulting $\bfv^\varepsilon$ is feasible and differs from $\bfv^\star$ by at most $(|\Rset_w|-1)\varepsilon$ in each coordinate of OD pair $w$. Thus $\bfv^\varepsilon\to\bfv^\star$, including at boundary points, and $\varepsilon\log Z_F^\varepsilon\ge\varepsilon\log\Psi_\varepsilon(\bfv^\varepsilon)=-I_\star+o(1)$. Together,
\begin{equation}
    \varepsilon\log Z_F^\varepsilon
    \longrightarrow
    -I_\star,
    \label{eq:app-partition-limit}
\end{equation}
which identifies $I(\bfv)-I_\star$ as the normalized limiting rate.

Let $\mathcal N$ be any open neighborhood of $\bfv^\star$. Continuity of $I$
on the compact feasible set and uniqueness of the minimizer imply that
\[
    \xi
    :=
    \inf_{\bfv\in\mathcal F(\mathbf q)\setminus\mathcal N}
    I(\bfv)-I_\star
    >0.
\]
Hence, by \cref{eq:app-uniform-rate,eq:app-partition-limit},
\[
    \pi_F^\varepsilon(\mathcal N^c)
    \le
    \left|\mathcal F_\varepsilon(\mathbf q)\right|
    \exp\!\left\{-\frac{\xi+o(1)}{\varepsilon}\right\}
    \longrightarrow0,
\]
which establishes concentration at $\bfv^\star$.

\paragraph{Identification with deterministic logit SUE.}
Since each $c_a$ is nondecreasing,
$B$ is convex, and so is $B(\Delta\bfv)$ in $\bfv$. The entropy term is strictly convex on
each OD simplex, so the objective in
\cref{eq:logit-sue-program} has a unique minimizer, which is also the unique minimizer of $I$.

Suppose $f_{wr}^\star=0$. Since $q_w>0$, some route $s$ has $f_{ws}^\star>0$. Transfer a small amount $\delta>0$ from $s$ to $r$. Bounded route costs make the Beckmann term change by $O(\delta)$, while the entropy term changes by $\theta^{-1}\delta\log(\delta/q_w)+O(\delta)$. The objective therefore decreases for sufficiently small $\delta$, a contradiction. Thus $f_{wr}^\star>0$; the nonnegativity constraints are inactive, and only the demand constraints carry multipliers. With multipliers $\zeta_w$ for the demand constraints, and since $\partial B(\Delta\bfv)/\partial f_{wr}=C_{wr}(\Delta\bfv)$, the first-order condition at $\bfv^\star$ is
\begin{equation}
    C_{wr}(\bx^\star)
    +
    \frac{1}{\theta}
    \left[
        \log\!\left(\frac{f_{wr}^\star}{q_w}\right)+1
    \right]
    +
    \zeta_w
    =
    0.
\end{equation}
Hence $f_{wr}^\star/q_w$ is proportional to $\exp\{-\theta C_{wr}(\bx^\star)\}$ over $r\in\Rset_w$, and the constraint $\sum_{r}f_{wr}^\star=q_w$ fixes the constant. This is Equation~\eqref{eq:logit-sue-fixed-point}.

\paragraph{Means and the limit $\theta\to\infty$.}
Let $\bF^\varepsilon$ be the random physical path flow. Its support lies in the compact set $\mathcal F(\mathbf q)$, whose diameter we denote by $d_{\mathbf q}$. For every $\delta>0$,
\[
  \E\|\bF^\varepsilon-\bfv^\star\|
  \le\delta+d_{\mathbf q}\,\Pp\bigl(\|\bF^\varepsilon-\bfv^\star\|>\delta\bigr).
\]
Concentration, followed by $\delta\downarrow0$, gives convergence of the path-flow means. Continuity and linearity of $\Delta$ give concentration of physical link flows and convergence of their means to $\bx^\star$.

Now let $\theta\to\infty$ after the nonatomic limit, keeping physical demands, routes, and costs fixed. Write $\bfv_\theta^\star$ for the logit SUE minimizer and $B_\star=\min_{\bfv\in\mathcal F(\mathbf q)}B(\Delta\bfv)$. The entropy sum $\sum_{w,r}f_{wr}\log(f_{wr}/q_w)$ lies in $[-C_{\mathbf q},0]$, where $C_{\mathbf q}=\sum_wq_w\log|\Rset_w|$. Comparing the objective at $\bfv_\theta^\star$ with its value at a Beckmann minimizer gives
\[
  0\le B(\Delta\bfv_\theta^\star)-B_\star\le C_{\mathbf q}/\theta.
\]
Compactness and continuity imply that every accumulation point minimizes the Beckmann potential and hence satisfies the Wardrop conditions on the fixed route sets.

\subsection{Proof of \texorpdfstring{\cref{thm:layered}}{the block-layered law}}
\label{app:layered}

\begin{proof}
(a) Fix block $g$ and write $o=o_g$ and $\mathcal T=\mathcal T_o$. Every node $i\ne o$ of $\mathcal T$ has at least one incoming link, because it is reachable from $o$, and $o$ has none, because $\mathcal T$ is acyclic. Consider a labeled assignment of block-$g$ travelers to bush paths. Every traveler present at a node $i\ne o$, either because $i$ is its destination or because its route continues from $i$, entered $i$ through exactly one link in $\delta^-(i)$. The assignment therefore determines, at each node, a partition of the travelers present there according to their incoming link. Conversely, any collection of such node-level partitions determines a unique assignment. Starting from its destination and following the selected incoming links, each traveler traces a path that terminates at $o$, the only node without incoming links. By \cref{ass:bush}, that path belongs to $\Rset_w$. The two descriptions are thus in bijection.

Now fix $\bx^g\in\Xset_g$ and process the nodes of $\V_o\setminus\{o\}$ in reverse topological order. When node $i$ is processed, the counts on its outgoing links are already fixed, so the number of travelers present at $i$ equals $m_i^g(\bx^g)$ and depends only on $\bx^g$. The number of ways to partition these $m_i^g$ labeled travelers among the incoming links with prescribed counts $(x_a^g)_{a\in\delta^-(i)}$ is the multinomial coefficient $m_i^g!/\prod_{a\in\delta^-(i)}x_a^g!$, which is well defined by Equation~\eqref{eq:layer-conservation}. Multiplying over nodes gives Equation~\eqref{eq:layer-multiplicity}.

(b) Under $\pi_R$, the probability of a profile depends on $\bR$ only through $\bx(\bR)=\sum_g\bx^g(\bR)$. Hence
\[
  \pi_L\bigl((\bx^g)_g\bigr)
  =\frac{1}{Z_R}\exp\Bigl\{-\theta\,\Phi\Bigl(\sum_g\bx^g\Bigr)\Bigr\}\,
   \bigl|\{\bR\in\mathcal S:\bx^g(\bR)=\bx^g\ \text{for all } g\}\bigr|.
\]
Blocks consist of disjoint sets of travelers whose route choices are not constrained across blocks, so the count factorizes as $\prod_g\Omega_g(\bx^g)$ by part (a).

(c) Divide Equation~\eqref{eq:layered-law} by the marginal probability of $(\bx^h)_{h\ne g}$. The factor $\exp\{\theta\Phi(\bb^g)\}$ does not depend on $\bx^g$ and is absorbed into the normalization.
\end{proof}

\subsection{Large-population limit of the layer multiplicity}
\label{app:layer-limit}

In the flow-quantum scaling of \cref{subsec:large-population-sue}, keep the block index set and physical terminating demands $q_i^g$ fixed, with $n_i^{g,\varepsilon}=q_i^g/\varepsilon\in\Nzero$. For feasible physical layers $\bx^{g,\varepsilon}\to\bx^g$, let $m_i^g=q_i^g+\sum_{a\in\delta^+(i)}x_a^g$, and define $\Omega_g^\varepsilon(\bx)=\Omega_g(\bx/\varepsilon)$ using Equation~\eqref{eq:layer-multiplicity} with demands $(n_i^{g,\varepsilon})_i$. Stirling's formula and Equation~\eqref{eq:layer-conservation} give
\[
  -\varepsilon\log\Omega_g^\varepsilon(\bx^{g,\varepsilon})
  \longrightarrow
  \sum_{i\in\V_{o_g}\setminus\{o_g\}}
  \Bigl[\,\sum_{a\in\delta^-(i)}x_a^g\log x_a^g
        -m_i^g\log m_i^g\Bigr],
\]
in physical flow units. The right-hand side is the negative demand-weighted path-choice entropy of the backward Markov decomposition with predecessor proportions $x_a^g/m_i^g$ at nodes of positive inflow, as in \citet{Akamatsu1997}. Under the cost assumptions of \cref{thm:large-population-sue}, applying this limit to each block--destination group as a commodity and then aggregating its path flows gives concentration at the corresponding logit SUE layers.

\subsection{Proof of \texorpdfstring{\cref{prop:surrogate}}{the frozen-cost surrogate}}
\label{app:surrogate}

\begin{proof}
(a) Under independent frozen-cost logit choices on bush routes, a block-$g$ traveler with destination $d$ uses route $r$ with probability $\exp\{-\theta C_{wr}(\by)\}/H_d(\by)$. A labeled assignment that produces the layer $\bz$ therefore has probability
\[
  \prod_{(w,j)\in g}\frac{\exp\{-\theta C_{wR_{wj}}(\by)\}}{H_{d_w}(\by)}
  =\frac{\exp\{-\theta\,\mathbf c(\by)^{\trans}\bz\}}{\prod_{i\in\V_g}H_i(\by)^{n_i^g}},
\]
because the route costs of all block travelers add up to $\sum_a z_a c_a(y_a)$. This probability depends on the assignment only through $\bz$, and by \cref{thm:layered}(a) exactly $\Omega_g(\bz)$ assignments produce $\bz$.

(b) The loading produces only layers in $\Xset_g$, since $\sum_{a\in\delta^-(i)}Z_a=m_i^g(\bZ)$ at every non-origin node $i\in\V_g\setminus\{o_g\}$. A given $\bz\in\Xset_g$ is produced with probability
\[
  \prod_{i\ne o_g}\left[\frac{m_i^g(\bz)!}{\prod_{a\in\delta^-(i)}z_a!}\prod_{a\in\delta^-(i)}\gamma_a(\by)^{z_a}\right]
  =\Omega_g(\bz)\prod_{a\in\A_g}\gamma_a(\by)^{z_a}.
\]
For $a=(j,i)$, $\gamma_a=H_j\omega_a/H_i$. Hence $\prod_a\gamma_a^{z_a}=\prod_a\omega_a^{z_a}\prod_{k\in\V_g}H_k^{e_k}$, where $e_k$ is the flow out of $k$ minus the flow into $k$. By Equation~\eqref{eq:layer-conservation}, $e_k=-n_k^g$ for $k\ne o_g$, and $H_{o_g}=1$. Since $\prod_a\omega_a^{z_a}=\exp\{-\theta\,\mathbf c(\by)^{\trans}\bz\}$, the probability equals Equation~\eqref{eq:surrogate-law}. Each node is processed once and each link receives one count, which gives the complexity statement.
\end{proof}

\subsection{Proof of \texorpdfstring{\cref{prop:block-mh}}{block-sampler exactness}}
\label{app:block-mh}

\begin{proof}
Fix the other layers, so that $\bb^g$ and $\by=\by(\bb^g)$ are fixed. The proposal $\widetilde\pi_g(\cdot\given\by)$ does not depend on the current layer $\bx^g$. By Equation~\eqref{eq:tilt}, the Hastings ratio for target $\pi_g(\cdot\given\bb^g)$ is
\[
  \frac{\pi_g(\bz'\given\bb^g)\,\widetilde\pi_g(\bx^g\given\by)}{\pi_g(\bx^g\given\bb^g)\,\widetilde\pi_g(\bz'\given\by)}
  =\exp\bigl\{-\theta\bigl[\rho_g(\bz';\bb^g,\by)-\rho_g(\bx^g;\bb^g,\by)\bigr]\bigr\},
\]
which gives Equation~\eqref{eq:mh-acceptance}. The block update is therefore a Metropolis--Hastings kernel with invariant distribution $\pi_g(\cdot\given\bb^g)$. By \cref{thm:layered}(c), this is the full conditional of block $g$ under $\pi_L$, so the update leaves $\pi_L$ invariant \citep{Tierney1994}, and so does any composition of block updates in a state-independent order. Because $\Omega_g>0$ on $\Xset_g$ and all node weights are positive, $\widetilde\pi_g(\cdot\given\by)$ is positive on $\Xset_g$ and every acceptance probability is positive. One sweep thus moves any state in $\prod_g\Xset_g$ to any other with positive probability, and the current state is retained with positive probability. The chain is irreducible and aperiodic, $\pi_L$ is its unique invariant distribution, and $\bx=\sum_g\bx^g$ has the pushforward $\pi_X$ in stationarity by \cref{thm:layered}(b).
\end{proof}

\subsection{Proof of \texorpdfstring{\cref{prop:block-error-bound}}{the rejection-rate bound}}
\label{app:block-error-bound}

\begin{proof}
Write $\mu=\pi_g(\cdot\given\bb)$ and $\nu=\widetilde\pi_g(\cdot\given\by)$. For $\bx,\bz\in\Xset_g$, let $A(\bx,\bz)=\mu(\bx)\nu(\bz)$ and $A^\leftarrow(\bx,\bz)=\nu(\bx)\mu(\bz)$. The mean acceptance $\bar\alpha_g$ averages over the independent current and proposed layers with joint law $A$. The Metropolis--Hastings rule gives
\[
  A(\bx,\bz)\alpha_g(\bx,\bz)=\min\{A(\bx,\bz),A^\leftarrow(\bx,\bz)\}.
\]
Since both joint laws have mass one,
\[
  1-\bar\alpha_g
  =\frac12\sum_{\bx,\bz}|A(\bx,\bz)-A^\leftarrow(\bx,\bz)|
  =\|\mu\otimes\nu-\nu\otimes\mu\|_{\mathrm{TV}}.
\]
Taking the first-coordinate marginal gives the lower bound $\|\mu-\nu\|_{\mathrm{TV}}$. The triangle inequality through $\nu\otimes\nu$ gives
\[
  \|\mu\otimes\nu-\nu\otimes\mu\|_{\mathrm{TV}}
  \le\|\mu\otimes\nu-\nu\otimes\nu\|_{\mathrm{TV}}
     +\|\nu\otimes\nu-\nu\otimes\mu\|_{\mathrm{TV}}
  =2\|\mu-\nu\|_{\mathrm{TV}}.
\]
At joint stationarity, each intermediate block update preserves $\pi_L$, so conditioning on its background and averaging yields the diagnostic in \cref{subsec:mh}.
\end{proof}%

\section{Implementation Details}
\label{app:implementation}

\paragraph{Demand integerization.}
At $\varepsilon=1$, Chicago Sketch demand is integerized separately within each origin. OD demands are first rounded down, and the remaining units needed to match the nearest integer origin total are assigned in descending order of the fractional remainders, with destination node number breaking ties. This procedure preserves a total of 1,137,496 travelers and leaves 386 of the 387 zones with positive interzonal demand. In the Sioux Falls quantum scan with $\varepsilon>100$, we round $q_w/\varepsilon$ down and allocate the remaining units in descending order of fractional remainder until the nearest integer network total is reached, breaking ties by origin and destination indices. This gives 1,803, 721, and 361 decision units at $\varepsilon=200$, 500, and 1000, respectively.

\paragraph{Additional warm-up.}
In the Sioux Falls precision scan, the runs with $\theta=0.7$ and $0.8$ at $\varepsilon=100$ use an additional 300 MH warm-up sweeps with $K=30$ before the discarded sweeps; all retained sweeps use $K=3$.

\paragraph{Software and hardware.}
The implementation uses Python 3.13.9, NumPy 2.3.5, SciPy 1.16.3, and Numba 0.62.1. Timings were obtained under Windows 11 on a workstation with an Intel Core i9 (Raptor Lake) processor with 32 logical processors; no more than four chains were run concurrently.

\section{Notation Summary}
\label{app:notation}

This appendix collects the notation used in the model and sampler.

{
\small
\begin{longtable}{@{}p{0.24\textwidth}p{0.72\textwidth}@{}}
\toprule
Symbol & Meaning \\
\midrule
\endfirsthead
\toprule
Symbol & Meaning \\
\midrule
\endhead
\bottomrule
\endfoot
\multicolumn{2}{@{}l}{\emph{Network, travelers, and costs}}\\
$G=(\V,\A)$ & Directed network with node set $\V$ and link set $\A$ \\
$\W$; $w=(o_w,d_w)$ & Set of OD pairs; OD pair with origin $o_w$ and destination $d_w$ \\
$n_w$; $N$ & Number of unit travelers of OD pair $w$; total number of travelers \\
$\Rset_w$ & Fixed finite nonempty route set of OD pair $w$ \\
$R_{wj}$; $\bR$; $\mathcal S$ & Route of traveler $(w,j)$; labeled route profile; route-profile space $\prod_w\Rset_w^{n_w}$ \\
$\Delta$; $\Delta_r$ & Link--route incidence matrix; its column for route $r$ \\
$\bx(\bR)$; $\bx^{-wj}$ & Link flow of a profile; link flow of all travelers except $(w,j)$ \\
$c_a$; $\mathbf c(\bx)$ & Cost function of link $a$; vector of link costs \\
$C_{wr}(\bx)$ & Cost of route $r$ of OD pair $w$ \\
$\theta$ & Common logit route-choice precision \\[2pt]
\multicolumn{2}{@{}l}{\emph{Potentials and scaling}}\\
$\phi_a$; $\Phi$ & Left-endpoint discrete link potential; network potential \\
$B$ & Beckmann potential \\
$\varepsilon$; $\Phi_\varepsilon$; $q_w$ & Flow quantum; discrete potential for flow quantum $\varepsilon$; physical OD demand \\
$f_{wr}^\varepsilon$; $N_{wr}^\varepsilon$ & Physical path flow; route count, with $f_{wr}^\varepsilon=\varepsilon N_{wr}^\varepsilon$ \\[2pt]
\multicolumn{2}{@{}l}{\emph{Exact laws}}\\
$\pi_R$; $Z_R$ & Joint Gibbs law of route profiles; common normalizer of the route, path, layered, and total-link laws \\
$\bF$; $\bfv$; $\Fset_n$ & Path counts; a path-count vector; feasible integer path-count set \\
$M(\bfv)$; $\pi_F$ & Path-count multiplicity; path-count law \\
$\mathcal F(\mathbf q)$ & Feasible physical path-flow set with demand $\mathbf q$ \\
$I$; $I_\star$ & Large-population objective; its minimum; $I-I_\star$ is the normalized limiting rate \\
$\bX$; $\bx$ & Random total link flow; its realization \\
$\Xset$; $\Dset(\bx)$ & Feasible integer link-flow set; fiber of path counts producing $\bx$ \\
$\Gamma(\bx)$; $\pi_X$ & Link-flow multiplicity; total link-flow law \\[2pt]
\multicolumn{2}{@{}l}{\emph{Bushes and block layers}}\\
$\mathcal T_o=(\V_o,\A_o)$ & Acyclic bush of origin $o$ (\cref{ass:bush}) \\
$\mathcal G$; $o_g$; $\mathcal T_g=(\V_g,\A_g)$ & Block partition of travelers; origin of block $g$; its bush \\
$n_i^g$ & Number of block-$g$ travelers with destination $i$ \\
$\bX^g$; $\bx^g$; $\Xset_g$ & Random block layer; its realization; feasible integer layer set \\
$\delta^-(i)$; $\delta^+(i)$ & Incoming and outgoing links of node $i$ in the bush \\
$m_i^g$ & Block-$g$ flow reaching a non-origin node $i$, equal to terminating demand plus outgoing flow \\
$\Omega_g$; $\pi_L$ & Layer multiplicity; joint law of the layers \\
$\bb^g$ & Leave-block-out background flow \\[2pt]
\multicolumn{2}{@{}l}{\emph{Block sampler}}\\
$\pi_g(\cdot\given\bb)$ & Exact block conditional \\
$\by$; $\hat\bz(\bb)$; $K$; $\eta_k$ & Reference flow; block reference iterate in $\operatorname{conv}(\Xset_g)$; fixed iteration count and damping values \\
$\omega_a(\by)$; $H_i(\by)$; $\gamma_a(\by)$ & Frozen-cost loading weight; node weight; predecessor probability \\
$\widetilde\pi_g(\cdot\given\by)$ & Frozen-cost surrogate (proposal) \\
$\bZ$; $\bz$ & Proposed block layer; a candidate layer \\
$\rho_g(\bz;\bb,\by)$ & Frozen-cost residual \\
$\alpha_g$ & Acceptance probability \\
$\bar\alpha_g$ & Mean acceptance at conditional stationarity for fixed background and reference, including accepted self-proposals \\
$U_g$; $\Sigma_g$; $D_g$ & Orthonormal basis for feasible-layer differences; proposal covariance and link-cost derivative matrix in those coordinates \\
$\lambda_k$ & Eigenvalues of $\theta\Sigma_g^{1/2}D_g\Sigma_g^{1/2}$ in the local Gaussian comparison (\cref{rem:acceptance-scaling}) \\
\end{longtable}
}

\end{document}